\documentclass[english,aps,prb,twocolumn,showpacs,floatfix]{revtex4}
\usepackage[T1]{fontenc}
\usepackage[latin1]{inputenc}
\usepackage{amsmath}
\usepackage{graphicx}

\usepackage{amssymb}

\makeatletter

\usepackage{amsfonts}

\usepackage{epsfig}
\usepackage{dsfont}

\newcommand{\Eq}[1]{Eq.~(\ref{#1})}
\newcommand{\Eqs}[1]{Eqs.~(\ref{#1})}

\newcommand{\be}{\begin{equation}}
\newcommand{\bea}{\begin{eqnarray}}
\newcommand{\eea}{\end{eqnarray}}
\newcommand{\ee}{\end{equation}}

\newcommand{\qqph}{\qquad \phantom{.}}

\newcommand{\NRG}{{{\rm NRG}}}

\newcommand{\VMPS}{{{\rm VMPS}}}

\newcommand{\Kept}{{{\rm K}}}
\newcommand{\Discarded}{{{\rm D}}}
\newcommand{\Total}{{{\rm A}}}
\newcommand{\XX}{{{\rm X}}}
\newcommand{\GG}{{{\rm G}}}

\newcommand{\pdagger}{{\phantom{\dagger}}}

\def\H{{\cal H}}

\newcommand{\SSS}{{\cal S}}
\newcommand{\UUU}{{\cal U}}
\newcommand{\VVV}{{\cal V}}
\newcommand{\EPsi}{E}
\newcommand{\DoS}{\rho_n}

\newcommand{\AP}{B}

\newcommand{\folded}{{\rm f}}
\newcommand{\unfolded}{{\rm u}}
\newcommand{\refolded}{{\rm r}}
\newcommand{\cloned}{{\rm c}}
\newcommand{\ground}{{\rm G}}

\newcommand{\kem}{{k_{\rm em}}}
\newcommand{\reference}{{\rm ref}}

\def\one{\ensuremath{\hbox{$\mathrm I$\kern-.6em$\mathrm 1$}}}

\newcommand{\DNRG}{{D}}
\newcommand{\DDMRG}{D'}
\newcommand{\dNRG}{{d}}
\newcommand{\dDMRG}{{d'}}
\newcommand{\dINRG}{{d_0}}

\newcommand{\bsigma}{{\boldsymbol{\sigma}}}

\newcommand{\VMPSgroundstate}{|\EPsi^N_\ground \rangle_\unfolded}

\newcommand{\mathone}{ 1 \hspace{-1.3mm} 1 }

\usepackage{babel}
\makeatother
\begin{document}

\title{Matrix product state comparison of the numerical
  renormalization group and the variational
formulation of the density matrix renormalization group}

\author{Hamed \surname {Saberi}}

\author{Andreas \surname{Weichselbaum}}

\author{Jan \surname{von Delft}}

\affiliation{Physics Department, Arnold Sommerfeld Center for Theoretical Physics,
and Center for NanoScience, Ludwig-Maximilians-Universität München,
80333 München, Germany}

\date{March 25, 2008}

\begin{abstract}
  Wilson's numerical renormalization group (NRG) method for solving
  quantum impurity models yields a set of energy eigenstates that 
  have the form of matrix product states (MPS).
  White's density matrix renormalization group (DMRG) for
  treating quantum lattice problems can likewise be reformulated in
  terms of MPS. Thus, the latter constitute a common algebraic
  structure for both approaches.  We exploit this fact to compare the
  NRG approach for the single-impurity Anderson model to 
  a variational matrix product state approach (VMPS),
  equivalent to single-site DMRG. 
  For the latter, we use an ``unfolded'' Wilson chain, which
  brings about a significant reduction in numerical costs compared
to those of NRG.  We show
  that all NRG eigenstates (kept and discarded) can be reproduced
  using VMPS, and compare the difference in truncation criteria, sharp
  vs. smooth in energy space, of the two approaches.  Finally, we
  demonstrate that NRG results can be improved upon systematically by
  performing a variational optimization in the space of variational
  matrix product states, using the states produced by NRG as input.
\end{abstract}



\maketitle

\section{Introduction}
\label{sec:intro}

Wilson's numerical renormalization group (NRG) is a highly successful
method for solving quantum impurity models which allows the
non-perturbative calculation of static and dynamic properties for a
variety of impurity models.\cite{Wilson,Krishna,Bulla,%
  AndersSchiller,AndersPruschkePRB06,WV} NRG is formulated on a
{}``Wilson chain'', i.e.\ a tight-binding fermionic quantum chain with
hopping matrix elements that decrease exponentially along the chain
as $\Lambda^{-n/2}$, where $\Lambda > 1$ is a discretization 
parameter defined below and $n \ge 0$ is the chain's
site index. It is thus not applicable to real space quantum lattice
problems featuring constant hopping matrix elements.  For these,
White's density matrix renormalization group (DMRG) is the method of
the choice.\cite{White,White2,Schollwoeck} It has been known for some
time\cite{Ostlund_Rommer,Dukelsky} that the approximate ground states
produced by DMRG have the form of matrix product states (MPS) (see
\Eq{eq:NRG_MPS} below) that had previously arisen in certain
stochastic models\cite{stoch_MPS} and quantum information
processing.\cite{AKLT+Fannes} This fact can be exploited to
reinterpret the DMRG algorithm (more precisely, its one-site
finite-size version) as a variational optimization scheme, in which
the ground state energy is minimized in the space of all matrix
product states with specified dimensions.\cite{VPC} To emphasize this
fact, we shall refer to DMRG as ``variational matrix product state''
(VMPS) approach throughout this paper.

Quite recently it was understood\cite{VMPS} that NRG, too, in a
natural way produces matrix product states. In other words, when
applied to the same Wilson chain, NRG and VMPS produce approximate ground states
of essentially the same MPS structure. The two approximate ground states are not
identical, though, since the two methods use different truncation
schemes to keep the size of the matrices involved manageable even for
very long Wilson chains: NRG truncation relies on energy scale
separation, which amounts to discarding the highest-energy eigenstates
of a sequence of effective Hamiltonians, say ${\cal H}_n$, describing
Wilson chains of increasing length $n$ and yielding spectral
information associated with the energy scale $\Lambda^{-n/2}$. This
truncation procedure relies on the exponential decrease of hopping
matrix elements along the Wilson chain, which ensures that adding
a new site to the Wilson chain perturbs it only weakly.  In contrast,
VMPS truncation relies on singular value decomposition of the matrices
consituting the MPS, which amounts to discarding the lowest-weight
eigenstates of a sequence of reduced density matrices.\cite{White2}
This procedure makes no special demands on the hopping matrix
elements, and indeed works also if they are all equal, as is 
the case of standard quantum chain models for which DMRG was designed.

The fact that a \emph{Wilson chain} model can be treated by two
related but inequivalent methods immediately raises an interesting and
fundamental methodological question: How do the two methods compare? More
precisely, to what extent and under which circumstances do their results
agree or disagree? How do the differences in truncation schemes
manifest themselves? VMPS, being a variational method operating in the
same space of states as NRG, will yield a lower-energy ground state
than NRG.  However, it variationally targets \emph{only} the ground
state for the full Wilson chain, of length $N$, say. In contrast, NRG
produces a set of eigenenergies $\{ E^n_\beta \} $ and eigenstates $\{
|E^n_\beta \rangle \} $ for each of the sequence of effective
Hamiltonians ${\cal H}_n$, with $n \le N$, mentioned above. From
these, a wealth of information about the RG flow, fixed points,
relevant and irrelevant operators, their scaling dimensions, as well
as static and dynamic physical properties can be extracted.  Are these
accessible to VMPS, too?

The goal of this paper is to explore such questions.  We shall exploit
the common matrix product state structure of the NRG and VMPS
approaches to perform a systematic comparison of these two methods, as
applied to the single-impurity Anderson model. It should be emphasized
that our purpose is not to advocate using one method instead of the
other. Instead, we hope to arrive at a balanced assessment of the
respective strengths and weaknesses of each method.

In a nutshell, the main conclusion (which confirms and extends the
results of Ref.~\onlinecite{VMPS}) is the following: when applied to a
Wilson chain with exponentially decreasing hopping, the VMPS approach
is able to fully reproduce \emph{all} information obtainable from NRG,
despite being variationally optimized with respect to the ground state
only. The reason is that the VMPS ground state is characterized by
products of matrices of the form $\prod_{n = 0}^N B^{[\sigma_n]}$
(details will be explained below), where the matrices with the same
index $n$ contain information about the energy scale $\Lambda^{-n/2}$.
As will be shown below, this information can be used to construct
eigenenergies $\{ E^n_\beta \} $ and eigenstates $\{ |E^n_\beta
\rangle \} $ for a sequence of effective Hamiltonians ${\cal H}_n$ in
\emph{complete} analogy with (but not identical to) those of NRG. 
The agreement between NRG and VMPS
results for these eigenenergies and eigenstates is excellent
quantitatively, provided sufficient memory resources are used for both
(and $\Lambda$ is not too close to 1, see below).  In this sense, NRG
and VMPS can be viewed as yielding essentially equivalent results when
applied to Wilson chains amenable to NRG treatment. In particular,
\emph{all} physical properties obtainable from the eigenspectra and
eigenstates of NRG can likewise be obtained from those of VMPS.

Nevertheless, NRG and VMPS do differ in performance, flexibility and
numerical cost.  Firstly, since NRG truncation relies on energy scale
separation, it works well only if the discretization parameter
$\Lambda$ is not too close to 1 (although the continuum limit of the
model is recovered only in the limit $\Lambda \to 1$). This
restriction does not apply to VMPS. Indeed, we shall find that NRG and
VMPS agree well for $\Lambda = 2.5$, but less well for $\Lambda =
1.5$. This in itself is not surprising. However it does illustrate the
power of VMPS to get by without energy scale separation. This very
useful feature can be exploited, for example, to obtain resolve sharp
spectral features at high energies in dynamical correlation
functions,\cite{unpublished} using projection operator
techniques. However, the latter results go beyond the scope of the
present paper and will be published separately.

Secondly, since VMPS does not rely on energy scale separation, it does
not need to treat all terms in the Hamiltonian characterized by the
same scale $\Lambda^{-n/2}$ at the same time, as is required for NRG.
This allows VMPS to achieve a significant reduction in memory cost
compared to NRG for representing the ground state.  To be specific:
For NRG, we use the standard ``folded'' representation of the Wilson
chain, in which each site represents both spin down and spin up
electrons, with the impurity site at one end (see
Fig.~\ref{fig_folded_unfolded}(a) below).  However, it turns out that
apart from the first few sites of the folded chain, the spin-down and
-up degrees of freedom of each site are effectively not entangled with
each other at all (see Fig.~\ref{fig_mutualinformation} below).  For
VMPS, we exploit this fact by using an ``unfolded'' representation of
the Wilson chain instead,\cite{Raas,VMPS} in which the spin up and
spin down sites lie on opposite sides of the impurity site, which sits
at the center of the chain (see Fig.~\ref{fig_folded_unfolded}(b)
below).  This unfolded representation greatly reduces the memory cost,
as characterized by the dimensions, $\DNRG$ for NRG or $\DDMRG$ for
VMPS, of the effective Hilbert spaces needed to capture the low energy
properties with the same precision: We find that with the choice
$\DDMRG=2^{m}\sqrt{\DNRG}$, VMPS can reproduce the results of NRG in
the following manner: (i) if $m=0$, the NRG ground state is reproduced
qualitatively; (ii) if $m=1$, all the ``kept'' states of NRG are
reproduced quantitatively; and (iii) if $m=2$ all the ``kept'' {\em
  and} ``discarded'' states of NRG are reproduced quantitatively.
However, in cases (ii) and (iii) the reduction in memory costs of VMPS
is somewhat offset by the fact that the calculation of the excited
eigenstates needed for the sake of direct comparison with NRG requires
diagonalizing matrices of effective dimension $D'^2$. Note,
nevertheless, that all information needed for this comparison is
already fully contained within the VMPS \emph{ground state}
characterized by dimension $\DDMRG$, since its constituent matrices
contain information from all energy scales represented by the Wilson
chain.

The paper is organized as follows: Section~\ref{sec:wilsonchains} sets
the scene by introducing a folded and an unfolded version of the
Wilson chain.  In Sections~\ref{sec:NRG_MPS} and~\ref{sec:DMRG_MPS} we
review the NRG and VMPS approaches for finding the ground state of a
folded or unfolded Wilson chain, respectively, emphasizing their
common matrix product state structure.  We also explain how an
unfolded MPS states may be ``refolded'', allowing it to be compared
directly to folded NRG states. In Section~\ref{sec:comparison} we
compare the results of NRG and VMPS, for ground state energies and
overlaps (Section~\ref{subsec:comparisonenergiesoverlaps}), excited
state eigenenergies and density of states
(Section~\ref{sec:eigenspectraDOS}), and the corresponding energy
eigenstates themselves (Section~\ref{sec:eigenstates}). This allows
us, in particular, to obtain very vivid insights into the differences
in the truncation criteria used by the NRG and VMPS approaches, being
sharp or smooth in energy space, respectively (Figs.~\ref{fig_DoS_2_5}
to \ref{fig_Wi_Ei2}).  In Section~\ref{sec:cloning+optimization} we
demonstrate that NRG results for the ground state can be improved upon
systematically by first producing an unfolded ``clone'' of a given NRG
ground state, and subsequently lowering its energy by performing
variational energy minimization sweeps in the space of variational
matrix product states. Finally, Section~\ref{sec:conclusions} contains our
conclusions and an assessment of the relative pros and cons of
NRG and VMPS in relation to each other.

\section{Folded and unfolded representations of  Wilson chain}
\label{sec:wilsonchains} 


For definiteness, we consider the single-impurity Anderson model.  It
describes a spinful fermionic impurity level with energy $\epsilon_d$
and double occupancy cost $U$ (with associated creation operators
$f^\dagger_{0 \mu}$, where $\mu = \downarrow , \uparrow$ denotes
spin), which acquires a level width $\Gamma$ due to being coupled to a
spinful fermionic bath with bandwidth $W=1$. Since the questions
studied in this paper are of a generic nature and do not depend much
on the specific parameter values used, we consider
only the symmetric Anderson model and take $U=\frac{1}{2}$,
$U/\pi\Gamma=1.013$ and $\epsilon_{d}=-\frac{1}{2}U$ throughout this
paper. To achieve a separation
of energy scales, following Wilson,\cite{Wilson,Krishna} the bath is
represented by a set of discrete energy levels with logarithmically
spaced energies $\Lambda^{-n}$ (with associated creation operators
$f^\dagger_{n \mu}$), where $n \ge 1$, $\Lambda>1$ is a
{}``discretization parameter'', and the limit $\Lambda \to 1$
reproduces a continuous bath spectrum.  The discretized Anderson model
Hamiltonian can then be represented as
 \begin{eqnarray}
  \H_{\mathrm{AM}}=\lim_{N\to\infty}\H_{N}\; ,
\label{eq:lim_SIAM}
\end{eqnarray}
where $\H_N$ describes a Wilson chain of ``length $N$''
(i.e., up to and including site $N$):
\begin{subequations}
  \label{eq:unfolded_SIAM1}
\begin{eqnarray} 
\H_N  & = & \H_{N \downarrow} + \H_{N \uparrow} + U
(f^\dagger_{0\uparrow} f^\pdagger_{0\uparrow}
 f^\dagger_{0\downarrow}  f^\pdagger_{0\downarrow} 
+{\textstyle \frac{1}{2}}) ,
\\
\H_{N \mu} & = & 
\epsilon_{d} f_{0\mu}^{\dagger}f^\pdagger_{0\mu}
+  \sum_{n=0}^{N-1}
  t_n (f_{n\mu}^{\dagger}  f^\pdagger_{(n+1)\mu} +\text{h.c.})  
\;,
\end{eqnarray}
\end{subequations}
%
with hopping coefficients given by
\begin{eqnarray}
t_{n} & \equiv & \begin{cases}
\sqrt{\frac{{2\Gamma}}{\pi}} & \text{for $n=0$} \; , \\
\frac{1}{2}(1+\Lambda^{-1})\Lambda^{-(n-1)/2 }\xi_{n} & 
\text{for $n \ge 1$} \; ,  \end{cases} \label{eq:Wilsoncouplings}
\rule[-5mm]{0mm}{0mm}
\\
\xi_{n} & = & 
(1-\Lambda^{-n})(1-\Lambda^{-2n+1})^{-1/2}
(1-\Lambda^{-2n-1})^{-1/2}\; . 
\nonumber \label{eq:xi_n}
\end{eqnarray}
In passing, we note that for our numerics we have found it convenient
(following Refs. \onlinecite{Raas} and \onlinecite{VMPS}) to keep
track of fermionic minus signs by making a Jordan-Wigner
transformation\cite{JW} of the Wilson chain to a spin chain, using
$f^\dagger_{n\mu} = P_{n \mu} s^+_{n\mu}$ and $f_{n\mu} = P_{n \mu}
s^-_{n\mu}$. Here $ s^\pm_{n\mu}$ are a set of spin-$\frac{1}{2}$
raising and lowering operators, that for equal indices satisfy $\{
s^-_{n\mu}, s^+_{n\mu} \} = 1$, $(s^-_{n\mu})^2 = (s^+_{n\mu})^2 = 0$,
but commute if their indices are unequal. The fermionic
anticommutation relations for the $f_{n\mu}$ are ensured by the
operators $P_{n\mu} = (-1)^{\sum_{(\bar n \bar \mu) < (n\mu)}
  s^+_{\bar n \bar \mu} s^-_{\bar n \bar \mu}}$, where $<$ refers to
some implicitly specified ordering for the composite index $(n\mu)$.
The $P_{n\mu}$ need to be kept track of when calculating certain
correlation functions, but do not arise explicitly in the construction
of the matrix product states that are the focus of this paper.  This
transformation will implicitly be assumed to have been implemented
throughout the ensuing discussion.

For the Anderson model, site $n$ of the Wilson chain represents the
set of four states $|\sigma_{n}\rangle $, with $\sigma_{n} =
(\sigma_{n \downarrow}, \sigma_{n \uparrow}) \in \{(00),(10),(01),(11)
\}$, where $\sigma_{n\mu} \in \{0,1\}$, to be viewed as eigenvalue of
$s^+_{n\mu} s^-_{n\mu}$, gives the occupancy on site $n$ of electrons
with spin $\mu$.  Thus, the dimension of the spinful index
$\sigma_{n}$ is $\dNRG=4$, and that of the spin-resolved index
$\sigma_{n\mu}$ is $\dDMRG = 2$.  As a general rule, we shall use the
absence or presence of primes, $\dNRG$ vs.\ $\dDMRG$ (and $\DNRG$ vs.\
$\DDMRG$ below), to distinguish dimensions referring to spinful or
spin-resolved indices, respectively, and correspondingly to folded or
unfolded representations of the Wilson chain. For other quantum
impurity models, such as the Kondo model or multilevel Anderson
models, the dimension of the local impurity site, say $\dINRG$,
differs from that of the bath sites, $\dINRG \neq \dNRG$. It is
straightforward to generalize the discussion below accordingly.

\begin{figure}[tb]
\centering 
\includegraphics[width=1\linewidth]{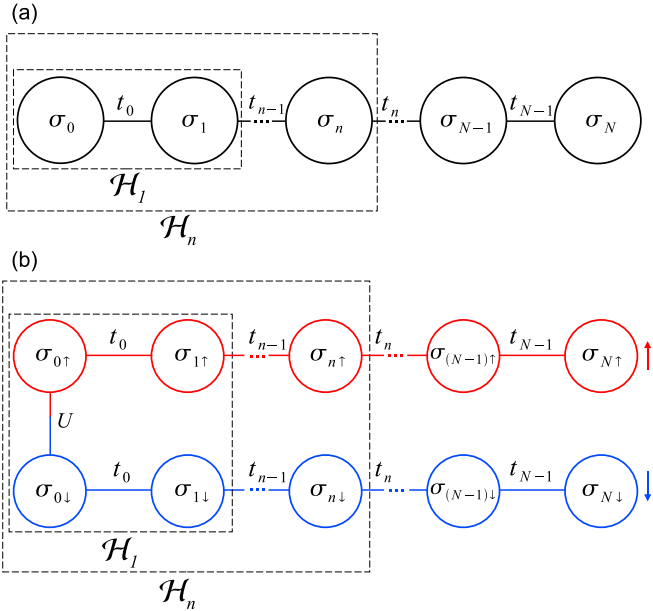} 
\caption{(Color online) (a) The standard
spinful or ``folded''  representation of the Wilson chain
of the single-impurity Anderson model, 
and (b) its spin-resolved or ``unfolded'' representation.
The latter makes explicit that
spin-down and -up states are coupled \emph{only} at the impurity sites
and not at any of the bath sites.
The dashed boxes indicate the chains 
described by $\H_1$ and $\H_n$, respectively.  
}
\label{fig_folded_unfolded}
\end{figure}

The Hamiltonian ${\cal H}_N$ of a Wilson chain of length $N$ is
defined on a Hilbert space of dimension $\dNRG^{N+1}$. It is spanned
by an orthonormal set of states that, writing $|\sigma_n \rangle =
|\sigma_{n\downarrow}\rangle |\sigma_{n \uparrow} \rangle$, can be
written in either spinful or spin-resolved form,
\begin{subequations}
\begin{eqnarray}
|\bsigma^N \rangle  & = & 
| \sigma_0 \rangle  
| \sigma_1 \rangle  \dots | \sigma_N \rangle, 
\label{eq:basissigma}
\\
& = & 
| \sigma_{0 \downarrow} \rangle
| \sigma_{0 \uparrow} \rangle | \sigma_{1 \downarrow} \rangle  
| \sigma_{1 \uparrow} \rangle  
\dots | \sigma_{N \downarrow} \rangle 
| \sigma_{N \uparrow} \rangle 
,   \qqph
\label{eq:basissigmaspinresolved}
\end{eqnarray}
\end{subequations}
corresponding to a ``folded'' or ``unfolded'' representation of the
Wilson chain, 
illustrated by Figs.~\ref{fig_folded_unfolded}(a) or (b), respectively.
The unfolded representation of Fig.~\ref{fig_folded_unfolded}(b) makes
explicit that the Anderson Hamiltonian of \Eq{eq:unfolded_SIAM1} has
the form
of two separate Wilson chains of specified spin, described by $\H_{N
  \downarrow}$ and $\H_{N \uparrow}$, \emph{which interact only at
  site zero}. This fact will be exploited extensively below.  Note
that the ordering chosen for the $|\sigma_{n\mu}\rangle$ states in
\Eq{eq:basissigmaspinresolved} fixes the structure of the many-body
Hilbert space once and for all.  The fact that the sites of the
unfolded chain in Fig.~\ref{fig_folded_unfolded} are connected in a
different order than that specified in \Eq{eq:basissigmaspinresolved}
is a statement about the dynamics of the model and of no consequence
at this stage, where we simply fix a basis.

\section{NRG treatment of folded Wilson chain}
\label{sec:NRG_MPS} 

\subsection{NRG matrix product state arises by iteration}
\label{sec:NRGiteration}

Wilson proposed to diagonalize the folded Wilson chain numerically
using an iterative procedure, starting from a short chain and adding
one site at a time.  Consider a chain of length $n$, sufficiently
short that $\H_n$ can be diagonalized exactly numerically.  Denote its
eigenstates by $|\EPsi_{\alpha}^{n}\rangle_\folded$, ordered by
increasing energy $(E_\alpha^{n})_\folded$, with $\alpha = 1 , \dots,
D_{n}$ and $D_n = d^{n+1}$. (We use subscripts $\folded$ and
$\unfolded$ to distinguish quantities obtained from a folded or
unfolded Wilson chain, respectively; similarly, in later parts of the
paper we will use the subscripts $\refolded$ and $\cloned$ for
``refolded'' and ``cloned''.)  E.g., for a chain consisting of only
the impurity site, $n=0$, the $\dNRG $ eigenstates can be written as
linear combinations of the form $|\EPsi^0_\alpha \rangle_\folded =
\sum_{\sigma_0} |\sigma_0 \rangle A^{[\sigma_0]}_{1\alpha} $, where
the coefficients have been arranged into $\dNRG $ matrices
$A^{[\sigma_0]}$ of dimensions $1 \times \dNRG$ (i.e.,
$\dNRG$-dimensional vectors), with matrix elements $A^{[\sigma_0]}_{1
  \alpha}$.  Then add to the chain the site $n+1$
and diagonalize $\mathcal{H}_{n+1}$ in the enlarged Hilbert space
spanned by the $(D_n \dNRG)$ states
$|\EPsi_{\alpha}^{n}\rangle_\folded|\sigma_{n+1}\rangle$.  The new
orthonormal set of eigenstates, with energies
$(E_\beta^{n+1})_\folded$, can be written as linear combinations of
the form
\begin{eqnarray}
  |\EPsi_{\beta}^{n+1}\rangle_\folded =\sum_{\sigma_{n+1}=1}^{\dNRG}
  \sum_{\alpha=1}^{D_n}|\EPsi_{\alpha}^{n}\rangle_\folded |\sigma_{n+1}\rangle\,
  A_{\alpha\beta}^{[\sigma_{n+1}]} \; ,
\label{eq:NRGrepresentation}
\end{eqnarray}
with $\beta=1,\dots,(D_{n} \dNRG)$. Here the coefficients specifying
the linear combination have been arranged into a set of $\dNRG$
matrices $A^{[\sigma_{n+1}]}$ of dimension $D_n \times D_{n+1}$,
with matrix elements $A_{\alpha\beta}^{[\sigma_{n+1}]}$.
The orthonormality of the eigenstates at each stage of the iteration,
${}_\folded \langle \EPsi_\beta^{n} |
\EPsi_{\beta'}^{n}\rangle_\folded = \delta_{\beta\beta'}$, implies
that the $A$-matrices automatically satisfy the orthonormality
condition
\begin{eqnarray}
  \sum_{\sigma_{n}}A^{[\sigma_{n}]\dagger}A^{[\sigma_{n}]}=\mathds{1}
\;. 
\label{eq:orthonormality}
\end{eqnarray}
We remark that it is possible to exploit symmetries of $\H_n$ (e.g.\
under particle-hole transformation) to cast $A$ in block-diagonal form
to make the calculation more time- and memory-efficient. However, for
the purposes of the present paper, this was not required.

Iterating the above procedure by adding site after site and repeatedly
using \Eq{eq:NRGrepresentation}, we readily find that the NRG
eigenstates of $\mathcal{H}_{N}$ on the folded Wilson chain can be
written in the form of a so-called \emph{matrix product
  state},\cite{VMPS}
\begin{eqnarray}
  |\EPsi^N_\beta\rangle_\folded =
\sum_{\lbrace\bsigma^N \rbrace}
|\bsigma^N \rangle  \,
(A^{[\sigma_0]} A^{[\sigma_1]} \dots A^{[\sigma_N]})_{1 \beta}
\; , 
\label{eq:NRG_MPS}
\end{eqnarray}
illustrated in Fig.~\ref{fig:matrix-product-structure}(a). Here matrix
multiplication is implied in the product, $(A^{[\sigma_n]}
A^{[\sigma_{n+1}]})_{\alpha \beta} = \sum_\gamma
A^{[\sigma_n]}_{\alpha \gamma} A^{[\sigma_{n+1}]}_{\gamma \beta}$, and
$\lbrace\bsigma^N \rbrace$ denotes the set of all sequences $\sigma_0,
\sigma_1, \dots, \sigma_N$.  This matrix multiplication generates
entanglement between neighboring sites, with the capacity for
entanglement increasing with the dimension $\DNRG_n$ of the index
being summed over.

\subsection{NRG truncation}
\label{sec:NRGtruncation}

\begin{figure}[tb]
\centering \includegraphics[width=1\linewidth]{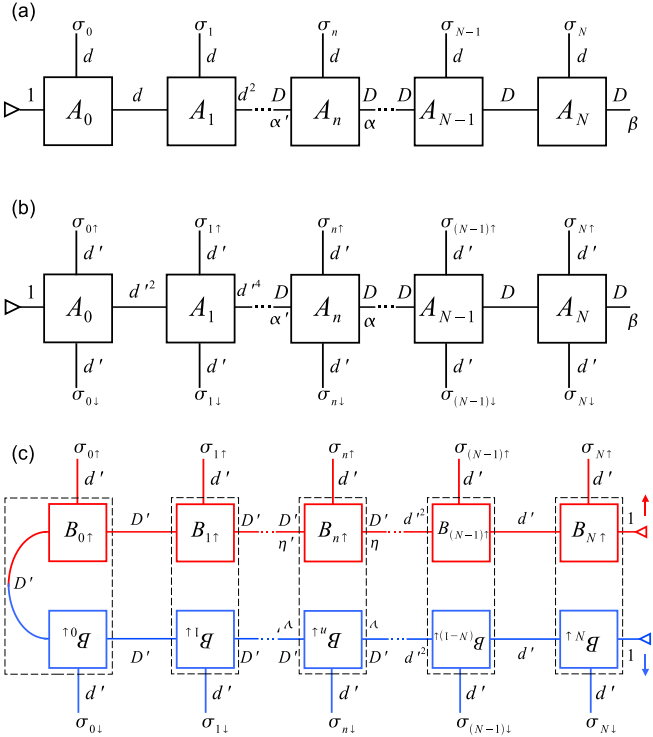} 
\caption{(Color online) 
  (a) and (b) show the matrix product structure of the state
  $|\EPsi^N_\beta\rangle_\folded$  of \Eq{eq:NRG_MPS}, depicting the site
  index as a single or composite index, $\sigma_n$ or
  $(\sigma_{n\downarrow}, \sigma_{n \uparrow})$, respectively. (c)
  shows the matrix product structure of the state
  $|\Psi^{N} \rangle_\unfolded$ of \Eq{eq:DMRG_MPS}.  (For the sake
  of illustrating \Eq{eq:BnBnBn} of Appendix~\ref{app:refolding}, the
  labels $(B_{n\downarrow})_{\nu \nu'}$ in the bottom row are
  purposefully typeset ``upside down'', so that they would be
  right-side up if the chain of boxes were all drawn in one row in the
  order indicated by \Eq{eq:DMRG_MPS}.)  Each matrix $A$ or $\AP$ is
  represented by a box, summed-over indices by links, free indices by
  terminals, and dummy indices having just a single value, namely 1,
  by ending in a triangle.  The dimensions ($\dNRG$, $\DNRG$,
  $\dDMRG$, $\DDMRG$, etc.)  next to each link or terminal give the
  number of possible values taken on by the corresponding index,
  assuming Wilsonian truncation for (a) and (b), and VMPS truncation
  for (c).  Note the similarity in structure between (c) and (b): the
  dashed boxes in the former, containing $\AP^{[\sigma_{n
      \downarrow}]}_{\nu \nu'} \otimes \AP^{[\sigma_{n
      \uparrow}]}_{\eta' \eta}$, play the role of the $A^{[(\sigma_{n
      \downarrow}, \sigma_{n \uparrow})]}_{\alpha' \alpha}$ matrices
  in the latter.  Their capacity for entangling neighboring sites is
  comparable if one chooses $\DDMRG^2 \propto \DNRG$ [cf.\
  Eq.~(\protect\ref{eq:DD'})], since neighboring dashed boxes in (c)
  are connected by two links of combined dimension $\DDMRG^2$, whereas
  neighboring $A$-matrices in (b) are connected by only a single link
  of dimension $\DNRG$. }
\label{fig:matrix-product-structure}
\end{figure}

In practice, it is of course not possible to carry out the above
iteration strategy explicitly for chains longer than a few sites,
because the size of the $A$-matrices grows exponentially with $N$.
Hence Wilson proposed the following NRG truncation procedure: Once
$D_n$ becomes larger than a specified value, say $\DNRG$, only the
lowest $D$ eigenstates $|\EPsi^n_\alpha \rangle_\folded$, with $\alpha = 1,
\dots, \DNRG$, are retained or kept at each iteration, and all
higher-lying ones discarded\cite{completebasis}.  Explicitly, the
upper limit for the sum over $\alpha$ in \Eq{eq:NRGrepresentation} is
redefined to be 
\begin{eqnarray}
D_n = \text{min} ( \dNRG^{n+1}, \DNRG) \; . 
\label{eq:NRG-dimensions}
\end{eqnarray}
As a result,
the dimensions of the $A^{[\sigma_n]}$ matrices occurring in the matrix
product state (\ref{eq:NRG_MPS}) 
start from $1\times \dNRG$ at $n=0$ and grow by a factor of $d$ for
each new site until they saturate at $\DNRG \times \DNRG$ after
truncation has set in.  The structure of the resulting states
$|\EPsi^N_\beta \rangle_\folded$ 
is schematically depicted in Figs.~\ref{fig:matrix-product-structure}(a)
and \ref{fig:matrix-product-structure}(b), in which the site index is
viewed as a single or composite index, $\sigma_n$ or
$(\sigma_{n\downarrow}, \sigma_{n\uparrow})$, respectively.

Wilson showed that this truncation procedure works well in practice,
because the hopping parameters $t_n$ of \Eq{eq:Wilsoncouplings}
decrease exponentially with $n$: the resulting separation of energy
scales along the chain ensures that high-lying eigenstates from
iteration $n$ make only a small contribution to the low-lying
eigenstates of iteration $n+1$, so that discarding the former hardly
affects the latter.  The output of the NRG algorithm is a set of
eigenstates $|\EPsi_\beta^n \rangle_\folded$ and eigenenergies
$(E_\beta^n)_\folded$ for each iteration, describing the physics at
energy scale $\Lambda^{-n/2}$.  The NRG eigenenergies are usually
plotted in rescaled form, 
\begin{equation}
\label{eq:NRGscaledeigenenergies}
(\varepsilon_\beta^n)_\folded = 
(E_\beta^n - E_1^n)_\folded / \Lambda^{- n/2} \; , 
\end{equation}
as functions of $n$, to obtain a so-called NRG
flow diagram; it converges to a set of fixed-point values as $n \to
\infty$. Figure~\ref{fig_flow_diagram} in Section~\ref{sec:flow_diag}
below shows some examples. The ground state energy of the
entire chain is given by the lowest energy of the last
iteration, $(E_G^N)_\folded = (E_1^N)_\folded$.

Despite the great success of NRG, Wilsonian truncation is does have some
drawbacks.  Firstly, its errors grow systematically as $\Lambda$ tends
to 1, because then the separation of energy scales on which it relies
becomes less efficient. Secondly, it is not variational, and hence it
is not guaranteed to produce the best possible approximation for the
ground state within the space of all matrix product states of similar
form and size.  We shall return to this point later in
Section~\ref{sec:cloning+optimization} and study quantitatively to
what extent the NRG ground state wavefunction can be improved upon by
further variational optimization.

\subsection{Mutual information of opposite spins on site $n$}
\label{sec:mutualinformation}

A crucial feature of the folded Wilson chain is that all degrees of
freedom associated with the same energy scale, $\Lambda^{-n/2}$, are
represented by one and the same site and hence are all added during
the same iteration step.  Since the spin-down and -up degrees of
freedom associated with each site are thus treated on an equal
footing, the resulting matrix product state provides comparable
amounts of resources for encoding entanglement between local states of
the same spin, involving $|\sigma_{n \mu} \rangle |\sigma_{n +1 \mu}
\rangle $, or between states of opposite spin (indicated by the bar),
involving $|\sigma_{n \mu} \rangle |\sigma_{n \bar \mu} \rangle $ or
$|\sigma_{n \mu} \rangle |\sigma_{n +1 \bar \mu} \rangle $.  However,
it turns out that for the Anderson model this feature, though \emph{a
  priori} attractive, is in fact an unnecessary (and memory-costly)
luxury: Since the Anderson model Hamiltonian (\ref{eq:unfolded_SIAM1})
couples spin-down and -up electrons only at the impurity site, the
amount of entanglement between states of opposite spin rapidly
decreases with $n$.


To illustrate and quantify this claim, it is instructive to calculate
the so-called mutual information $M^{\downarrow \uparrow}_n$ of the
spin-down and -up degrees of freedom of a given site $n$. 
This quantity is defined via the following general
construction.\cite{reducedDM} Let $C$ denote an arbitrary set of
degrees of freedom of the Wilson chain, represented by the states
$|\sigma^C_{\phantom{C}} \rangle$. Let $\rho^{C}$ be the reduced density matrix
obtained from  the ground state density matrix by tracing out all degrees of
freedom except those of $C$, denoted by $N\backslash C$:
\begin{eqnarray}
  \label{eq:rho_AB}
  \rho^{C} = \sum_{\{ \sigma^{N \backslash C} \}} 
 \langle  \sigma^{N \backslash C}  |E_G^N \rangle_\folded {}_\folded 
   \langle  E_G^N | \sigma^{N \backslash C} \rangle \; .
\end{eqnarray}
For example, if $C$ represents the spin-down and up-degrees
of freedom of site $n$,
its matrix elements are:
\begin{eqnarray}
\nonumber
  \rho^C_{\sigma_{n} \sigma'_{n}} & = & 
  \sum_{\lbrace\bsigma^{N \backslash n} \rbrace} 
  (A^{[\sigma_N]\dagger} \dots A^{[\sigma_n]\dagger} \dots
  A^{[\sigma_0]\dagger})_{\ground 1} 
\\ 
  \label{eq:rhosigman}
 & &  \qquad \times 
  (A^{[\sigma_0]} \dots A^{[\sigma'_n]} \dots 
  A^{[\sigma_N]})_{1 \ground }  \; .
\end{eqnarray}
If $C$ represents only the spin-$\mu$ degree of freedom of site $n$,
a similar expression holds, with $n$ replaced by $n \mu$.
The entropy associated with such a density matrix is given by
\begin{eqnarray}
  \label{eq:entropy}
  S^C = - \sum_i w_i^C \ln w_i^C \; ,
\end{eqnarray}
where $w_i^C$ are the eigenvalues of $\rho^C$, with
$\sum_i w_i^C = 1$.  Now, consider the case
that $C = AB$ is a combination of the degrees of freedom of two
distinct subsets $A$ and $B$, represented by states of the form
$|\sigma^C_{\phantom{C}} \rangle = |\sigma^A_{\phantom{C}} \rangle 
|\sigma^B_{\phantom{C}} \rangle$.  
Then the mutual information of $A$ and $B$, defined by
\begin{eqnarray}
  \label{eq:mutualinformation}
  M^{AB} = S^A + S^B - S^{AB} \; , 
\end{eqnarray}
characterizes the information contained in $\rho^{AB}$ beyond that
contained in $\rho^A \otimes \rho^B$.  The mutual information $M^{AB}
= 0$ if there is no entanglement between the degrees of freedom of $A$
and $B$, since then $\rho^{A B} = \rho^A \otimes \rho^B$ and its
eigenvalues have a product structure, $w^{AB}_{ij} = w^A_i w^B_j$.

\begin{figure}[!t]
 \includegraphics[width=1\linewidth]{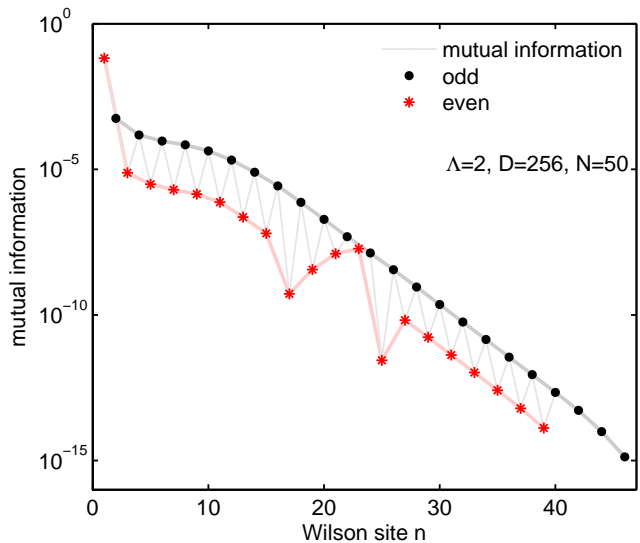} 
 \caption{NRG result for the mutual information $M_n^{\downarrow \uparrow}$
between spin-down and -up degrees of freedom of site $n$
of a folded Wilson chain of lenght $N=50$.  
 The Anderson model parameters are fixed at
   $U=\frac{1}{2}$, $U/\pi\Gamma=1.013$, $\epsilon_{d}=-\frac{1}{2}U$
   throughout this paper.  Lines connecting data points are
guides for the eye. The slight differences in 
behavior observed for even or odd $n$ are reminiscent
of the well-known fact\cite{Wilson} that the ground 
state degeneracy of a Wilson chain is different for even or odd $N$. }
\label{fig_mutualinformation}
\end{figure}

We define the mutual information between spin-down and -up degrees of
freedom of site $n$ of the folded chain, $M^{\downarrow \uparrow}_n$,
by \Eq{eq:mutualinformation}, taking $A = {n \! \downarrow} $ and $B=
{n \! \uparrow} $. Figure~\ref{fig_mutualinformation} shows this
quantity as function of $n$ for the symmetric Anderson
model. Evidently $M_n^{\downarrow \uparrow}$ is very small for all but
the first few sites, and decreases exponentially with $n$.  This
implies that for most of the folded chain, there is practically no
entanglement between the spin-down and -up degrees of freedom.
Consequently, the corresponding matrices occuring in \Eq{eq:NRG_MPS}
for $ |E_G^N \rangle_\folded $ in effect have a direct product
structure: loosely speaking, we may write $A^{[\sigma_{n}]} \simeq
\AP^{[\sigma_{n\downarrow}]} \otimes \AP^{[\sigma_{n \uparrow}]}$. In
the next subsection, we will exploit this fact to achieve a
significant reduction in memory cost, by implementing the effective
factorization in an alternative matrix product Ansatz [see \Eq
{eq:DMRG_MPS} below], defined on an \emph{unfolded} Wilson chain which
represents $n \! \downarrow$ and $n \! \uparrow$ of freedom by two
separate sites.

\section{DMRG treatment of unfolded Wilson chain}
\label{sec:DMRG_MPS} 

\subsection{Variational matrix product state Ansatz}
\label{sec:MPSAnsatz}

As pointed out by Verstraete \emph{et al.},\cite{VMPS} an alternative
approach for finding a numerical approximation for the ground state can be
obtained by \emph{variationally minimizing} the ground state energy
in the space of all ``variational matrix product states'' (VMPS) of
fixed norm. Implementing the latter constraint via a
Lagrange multiplier $\lambda$, one thus considers the following
minimization problem,
\begin{eqnarray}
  \min_{|\Psi \rangle \in \{ |\Psi^N \rangle_\unfolded \} } \left[
    \langle\Psi| \H_{N} |\Psi \rangle  - \lambda 
    (\langle \Psi | \Psi \rangle -1 ) \right] \; .
\label{eq:VMPS_recipe}
\end{eqnarray}
The minimization is to be performed over the space of all
variational matrix product states $|\Psi^N \rangle_\unfolded $ having
a specified structure (see below), with specified dimensions $D'_n$
for the matrices, whose matrix elements are now treated as variational
parameters.  This minimization can be performed by a ``sweeping
procedure'', which optimizes one matrix at a time while keeping all
others fixed, then optimizing the neighboring matrix, and so forth,
until convergence is achieved.  The resulting algorithm is equivalent
to a single-site DMRG treatment of the Wilson chain. 
Our main goal is to analyse how the energies and
eigenstates so obtained compare to those produced by NRG.

Having decided to use a variational approach, it becomes possible to
explore matrix product states having different, possibly more
memory-efficient structures than those of \Eq{eq:NRG_MPS} and
Fig.~\ref{fig:matrix-product-structure}(a).  In particular, we can
exploit\cite{Raas} the fact that the Anderson model Hamiltonian
(\ref{eq:unfolded_SIAM1}) couples spin-down and -up electrons 
only at the impurity site, as emphasized in \Eq{eq:unfolded_SIAM1} and
Fig.~\ref{fig_folded_unfolded}(b).  For such a geometry, it is natural to
consider matrix product states defined on the \emph{unfolded} Wilson
chain (subscript \unfolded) and having the following form, depicted
schematically in Fig.~\ref{fig:matrix-product-structure}(c):
\begin{eqnarray}
  |\Psi^{N} \rangle_\unfolded =
  \sum_{\lbrace\bsigma^N \rbrace} \! 
  |\bsigma^N\rangle
  (\AP^{[\sigma_{N\downarrow}]} \!
  \dots \AP^{[\sigma_{0\downarrow}]}
  \AP^{[\sigma_{0\uparrow}]} \!
  \dots \AP^{[\sigma_{N\uparrow}]})_{11}
  . \nonumber 
  \nonumber \hspace{-0.5cm} \phantom{.}
  \\
\label{eq:DMRG_MPS}
\end{eqnarray}
The order in which the $B^{[\sigma_{n\mu}]}$ matrices
occur in the product mimics the order in which the sites
are connected in the unfolded Wilson chain. (The fact that this
order differs from the order in which the basis states 
$|\sigma_{n\mu}\rangle$ for each site are arranged
in the many-body basis state $|\bsigma^N\rangle$, 
see \Eq{eq:basissigmaspinresolved}, does not 
cause minus signs complications, 
because we work with Jordan-Wigner-transformed
effective spin chains.)
Each $\AP^{[\sigma_{n \mu}]}$ stands for a set of $\dDMRG = 2$
matrices with matrix elements $\AP^{[\sigma_{n \mu}]}_{\nu \eta}$,
 with dimensions $\DDMRG_{n} \times \DDMRG_{n-1}$ for
$\AP^{[\sigma_{n \downarrow}]}$ and $\DDMRG_{n-1} \times \DDMRG_{n}$
for $\AP^{[\sigma_{n \uparrow}]}$, where 
\begin{equation}
  \label{eq:DDMRG_n}
\DDMRG_n = \mathrm{min}(\dDMRG^{N-n}, \DDMRG),
\end{equation}
as indicated on the links connecting the squares in
Fig.~\ref{fig:matrix-product-structure}(c).  This choice of matrix
dimension allows the outermost few sites at both ends of the unfolded
chain to be described exactly (similarly to the first few sites of the
folded Wilson chain for NRG), while introducing truncation, governed
by $\DDMRG$, for the matrices in the central part of the chain.  The
first index on $\AP^{[\sigma_{N\downarrow}]}_{1\nu}$ and the second
index on $\AP^{[\sigma_{N\uparrow}]}_{\nu 1}$ are dummy indices taking
on just a single value, namely 1, since they represent the ends of the
chain.  The triangles in Fig.~\ref{fig:matrix-product-structure}(c)
are meant to represent this fact. As a result, \Eq{eq:DMRG_MPS}
represents just a single state, namely the ground state, in contrast
to \Eq{eq:NRG_MPS}, which represents a set of states, labeled by the
index $\beta$. Moving inward from the endpoints by decreasing $n$, the
matrix dimension parameter $\DDMRG_n$ increases by one factor of $\dDMRG$
for each site, in such a way that the resulting matrices are of just
the right size to describe the outside ends of the chain (from $n$ to
$N$) \emph{exactly}, i.e.\ without truncation. After a few sites,
however, truncation sets in and the matrix dimensions saturate at
$\DDMRG \times \DDMRG$ for the central part of the chain.



To initialize the variational search for optimal $\AP$-matrices, it
turns out to be sufficient to start with a set of random matrices with
normally distributed random matrix elements. Next, singular value
decomposition 
is used to orthonormalize the $\AP$-matrices in such a way [see
\Eq{subeq:orthonormality}] that the matrix product state
\Eq{eq:DMRG_MPS} has norm 1 (see App.~\ref{sec:orthonormalization} for
details).  Thereafter, variational optimization sweeps are performed
to minimize \Eq{eq:VMPS_recipe} one $\AP$-matrix at a time
\cite{VMPS}. (The technical details of this procedure will be
published separately\cite{Muender}.)  After a sweeping back and forth
through the entire chain a few times, the variational state typically
converges (as illustrated by Fig.~\ref{fig_swoNRG} in
Sec.~\ref{sec:convergence-energyoptimization} below), provided that
$\DDMRG$ is sufficiently large. We shall denote the resulting
converged variational ground state by $\VMPSgroundstate$.  Its
variational energy, $(E^N_\ground )_\unfolded$, turns out to be
essentially independent of the random choice of initial matrices.

\subsection{VMPS truncation}
\label{sec:DMRGtruncation}

Since $\DDMRG \times \DDMRG$ is the maximal dimension of $\AP$-matrices,
$\DDMRG$ is the truncation parameter determining the effective size of the
variational space to be searched and hence the accuracy of the
results. Its role can be understood more explicitly using a technique
that is exceedingly useful in the VMPS approach, namely singular value
decomposition: any rectangular matrix ${\cal B}$ of dimension $m
\times m'$ can be written as
\begin{equation}
{\cal B}=\UUU \SSS \VVV^\dagger \; , \quad \textrm{with} \quad 
\UUU^\dagger \UUU =
\VVV^\dagger \VVV =\mathds{1}
 \; , 
\label{eq:SVD}
\end{equation}
where $\SSS$ is a diagonal matrix of dimension
$\min(m,m')$, whose diagonal elements, the so-called ``singular values'',
can always be chosen to be real and non-negative, and $\UUU$ and
$\VVV^\dagger$ are column- and row-unitary matrices (with dimensions $m
\times \min(m,m')$ and $\min(m,m') \times m'$, respectively).  Due to
the latter fact, the matrix norm of ${\cal B}$ is governed by the
magnitude of the singular values.

For any given site of the unfolded Wilson chain,
this decomposition can be applied in one of two ways (depending
on the context, see App.~\ref{app:A}) to the set of matrices
with elements $\AP^{[\sigma_{n\mu}]}_{\nu \eta}$:
introduce a composite index $\bar \nu = (\sigma_{n\mu}, \nu)$ (or 
$\bar \eta = (\sigma_{n\mu}, \eta)$) to arrange their matrix elements into a
rectangular matrix carrying only two labels, with matrix elements
${\cal B}_{\bar \nu \eta} = \AP^{[\sigma_{n\mu}]}_{\nu \eta}$ (or
$\tilde {\cal B}_{\nu \bar \eta} = \AP^{[\sigma_{n\mu}]}_{\nu
  \eta}$), and decompose this new matrix as ${\cal B} = \UUU \SSS
\VVV^\dagger$. 

\begin{figure}[tb]
\includegraphics[width=1.\linewidth]{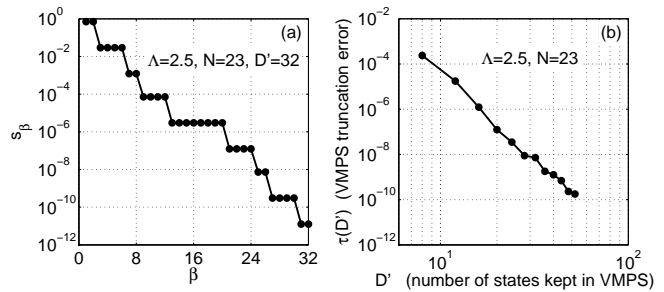} 
\caption{(Color online) (a) Typical singular value spectrum for site
  $5 \!\! \downarrow$ of the unfolded Wilson chain, obtained by
  singular value decomposition of $\AP^{[\sigma_{5\downarrow}]}$.
  It shows, roughly, power-law decrease for large enough $\beta$,
  modulo steps due to degeneracies in the singular value spectrum.
  (b) $\DDMRG$-dependence of the truncation error $ \tau (\DDMRG)$
  [\Eq{eq:truncationerror}]. }
\label{fig_singularvalues}
\end{figure}

Now, if this is done for any site for which the set of matrices
$\AP^{[\sigma_{n\mu}]}$ have maximal dimensions $\DDMRG \times
\DDMRG$, the corresponding matrix $\SSS$ will likewise have
dimensions $\DDMRG \times \DDMRG$. Let its diagonal elements, the
singular values $s_\nu$ (with $\nu = 1, \dots, \DDMRG$), be labelled
in order of decreasing size. (Their squares, $s_\nu^2$, correspond to
the eigenvalues of the density matrix constructed in the course of the
single-site DMRG algorithm\cite{White2}.)  If $\DDMRG$ is sufficiently
large, the $s_\nu$ are typically found to decrease with increasing
$\nu$ roughly as some negative power of $\nu$, as illustrated in
Fig.~\ref{fig_singularvalues}(a).  The last and smallest of the
singular values, $s^2_{\DDMRG}$ (squared, following
Ref.~\onlinecite{White2}), thus indicates the weight of the information
that is lost at that site due to the given (finite) choice of
$\DDMRG$: by choosing $\DDMRG$ larger, less information would be lost
since more singular values (though of smaller size) would be retained.
Repeating such an analysis for all sites of the unfolded Wilson chain,
one may define the largest of the $s^2_{\DDMRG}$ parameters of the
entire chain,
   \begin{equation}
     \label{eq:truncationerror}
     \tau(\DDMRG)=\max_{\{ n \mu \}} (s^2_{\DDMRG}) \; ,
   \end{equation}
   as ``truncation error'' characterizing the maximal information loss
   for a given value of $\DDMRG$.  Typically, $\tau (\DDMRG)$
   decreases as some negative power of $\DDMRG$, as illustrated in
   Fig.~\ref{fig_singularvalues}(b).  In this way, $\DDMRG$ assumes
   the role of a cutoff parameter that directly governs the accuracy
   of the VMPS approach, in a way analogous to the parameter $\DNRG$
   of NRG.

\subsection{Refolding}
\label{sec:refolding}

The VMPS approach purposefully focusses on finding an optimal
description of the variational \emph{ground} state $\VMPSgroundstate$.
Nevertheless, the $\AP$-matrices from which the latter is constructed
contain information about all energy scales of the model, due to the
logarithmic discretization of the Wilson chain. In particular,
information about the scale $\Lambda^{-n/2}$ is encoded in the set of
matrices $\AP^{[\sigma_{n\mu}]}$ associated with the two site $n \! \!
\downarrow$ and $n \! \! \uparrow$. From these, it is possible to
extract excited-state eigenspectra and energy flow diagrams in
complete analogy to those produced by NRG. In this subsection we
explain how this can be accomplished by a technique to be called
``refolding'', which combines the two matrices
$\AP^{[\sigma_{n\downarrow}]}$ and $\AP^{[\sigma_{n\uparrow}]}$ into a
single matrix, say $\AP^{[\sigma_n]}$, and thereby recasts unfolded
matrix product states into folded ones. It should be emphasized that
this procedure simply amounts to an internal reorganization of the
representation of the VMPS ground state.

Consider a given matrix product state $|\Psi^N \rangle_\unfolded$ of
the form (\ref{eq:DMRG_MPS}), defined on an unfolded Wilson chain of
length $N$ (e.g.\ the converged ground state $\VMPSgroundstate$).  To
\emph{refold} it (subscript r) , it is expressed as a state of
the following form
[same as \Eq{eq:NRG_MPS}]
\begin{eqnarray}
  |\Psi^N \rangle_\refolded =
\sum_{\lbrace\bsigma^N \rbrace}
|\bsigma^N \rangle
(\AP^{[\sigma_0]} \AP^{[\sigma_1]} \dots \AP^{[\sigma_N]})_{11}
\; ,
\label{eq:refolded_DMRG_MPS}
\end{eqnarray}
defined on a folded Wilson chain of length $N$ and normalized
to unity, 
${}_\refolded \langle \Psi^n |\Psi^n \rangle_\refolded = 1$. 
 Graphically speaking, this corresponds to
rewriting a state of the form shown in
Fig.~\ref{fig:matrix-product-structure}(c) in terms of states of the
form of Fig.~\ref{fig:matrix-product-structure}(a).  
To obtain the matrices needed for \Eq{eq:refolded_DMRG_MPS}, one
constructs, for every site $n$ of the refolded chain, a set of $\dNRG$
matrices $\AP^{[\sigma_n]}$ from a combination of the two sets of
spin-resolved matrices $\AP^{[\sigma_{n\downarrow}]}$ and
$\AP^{[\sigma_{n\uparrow}]}$ of the unfolded chain
(App.~\ref{app:refolding} gives the details of this construction).
This is done in such a way, using singular value decomposition, that
(i) the resulting matrices $\AP^{[\sigma_n]}$ satisfy the
orthonormality conditions (\ref{eq:orthonormality}) (with $A \to \AP$),
thereby guaranteeing the unit normalization of the
the refolded state $|\Psi^N \rangle_\refolded $; 
and (ii) the $\AP^{[\sigma_n]}$
matrices have a structure similar to that of the matrices
$A^{[\sigma_n]}$ generated by NRG, except that their dimensions,
$D^\refolded_n \times D^\refolded_{n+1}$, are governed by
\begin{equation}
  D^\refolded_n = \min(\dNRG^{n}, \dNRG^{N+1-n}, \DDMRG^2) 
  \label{eq:DNRG-refolded}
\end{equation}
[instead of
\Eq{eq:NRG-dimensions}], for reasons explained
in App.~\ref{app:refolding}. Thus, their dimensions 
have the maximal value $\DNRG^\refolded \times 
\DNRG^\refolded$, with $\DNRG^\refolded = \DDMRG^2$, 
in the central part of the refolded chain, while
decreasing at its ends towards 
$1\times \dNRG$ or $\dNRG \times 1$ for $n=0$ or $N$, respectively.

Now, suppose that a converged variational ground state
$\VMPSgroundstate$ has been obtained and refolded
into the form $ |\Psi^N\rangle_\refolded $, so that the
corresponding orthonormalized
matrices $\AP^{[\sigma_{n}]}$ for the refolded Wilson chain 
of length $N$ are the building blocks of the ground state
of the system.  Then it is possible to extract
from them information about the many-body excitation spectrum at
energy scale $\Lambda^{-n/2}$ that is analogous to the information
produced by NRG. To this end, consider a \emph{subchain} of length
$n$ of the full refolded Wilson chain, and use the definition
\begin{eqnarray}
  |\Psi^n_\beta \rangle_\refolded =
  \sum_{\lbrace\bsigma^n \rbrace}
  |\bsigma^n \rangle
  (\AP^{[\sigma_0]} \AP^{[\sigma_1]} \dots \AP^{[\sigma_n]})_{1 \beta}
  \; , 
\label{eq:refolded_DMRG_MPS_basis}
\end{eqnarray}
[as in \Eq{eq:refolded_DMRG_MPS}, but
with $N$ replaced by $n$] to construct a set of 
states $ |\Psi^n_\beta\rangle_\refolded $
on this subchain. 
These states, shown schematically by sites 0 to $n$ of
Fig.~\ref{fig:matrix-product-structure}(a), form an orthonormal set,
${}_\refolded \langle \Psi^n_\alpha |\Psi^n_\beta \rangle_\refolded =
\delta_{\alpha \beta}$, due to the orthonormality
[\Eq{eq:orthonormality}] of their constituent matrices.  They can thus
be viewed as a basis for that \emph{subspace} of the many-body
Hilbert space for the length-$n$ Wilson chain, i.e.\ of 
that subspace of ${\rm
  span}\{|\bsigma^n \rangle \}$, which VMPS sweeping has singled out
to be most relevant for describing the ground state $\VMPSgroundstate$
of the full chain of length $N$. Therefore we shall henceforth
 call the $ |\Psi^n_\beta \rangle_\refolded $ ``(refolded) VMPS
basis states'' for this subchain.

This basis can be used to define an effective ``refolded Hamiltonian''
$ \H^n_\refolded$ for this subchain, with matrix elements
\begin{equation}
  \label{eq:Heffective}
(  \H^n_\refolded)_{\alpha \beta} = 
{}_\refolded \langle \Psi^n_\alpha |
\H_n | \Psi^n_\beta\rangle_\refolded \; .
\end{equation}
Its eigenvalues and eigenstates, say $(E_{\beta}^n)_\refolded$ and
$|E_\beta^n\rangle_\refolded$,
are the VMPS analogues of the NRG eigenvalues and eigenstates,
$(E_\beta^n)_\folded$ and $|\EPsi_\beta^n\rangle_\folded$,
respectively. They differ, in general, because VMPS and NRG use
different truncation criteria, but are expected to agree well for
sufficiently large choices of $\DDMRG$ and $\DNRG$. This is indeed
found to be the case, as will be shown in detail in the next section.

\section{Comparison of NRG and VMPS results}
\label{sec:comparison}

Having outlined the NRG and VMPS strategies in the previous section,
we now turn to a comparison of their results.  This will be done, in
successive subsections, by comparing their ground state energies and
the overlaps of the corresponding ground states; the eigenspectra and
density of states obtained from both approaches; and finally, the
energy eigenstates used in the two approaches. We will thereby gain
more insights into the differences between NRG and VMPS truncation
criteria.  Before embarking on a detailed comparison, though, some
remarks on the choices to be made for $\DNRG$ and $\DDMRG$ are in
order.

Since the structure of the matrix products occurring in
\Eqs{eq:NRG_MPS} and (\ref{eq:DMRG_MPS}) differ, the spaces consisting
of all states of the type $|\EPsi^n_\beta\rangle_\folded$ or
$|E^n_\beta\rangle_\refolded$, to be called the ``NRG-subspace'' or
``VMPS-subspace'' for a length-$n$ chain, respectively, constitute
nonidentical subspaces of the $ \dNRG^{n+1}$-dimensional Hilbert space
spanned by the basis states $|\bsigma^n\rangle$.  The extent to which
they describe the energy eigenstates of ${\cal H}_N$ with comparable
accuracy will depend very strongly on the choices made for $\DNRG$ and
$\DDMRG$.  It turns out (numerical evidence will be presented below)
that with the choice
\begin{eqnarray}
\DDMRG = \dDMRG^m \sqrt{\DNRG} \; ,
\label{eq:DD'}
\end{eqnarray}
the VMPS-subspace is sufficiently large to give highly accurate
representations of all kept states of NRG (including, in particular,
the ground state) for the choice $m=1$, or of all kept \emph{and}
discarded states of NRG for the choice $m=2$.
The fact that $\DDMRG$ should be proportional to $\sqrt \DNRG$ can be
made plausible by considering the following question: given a folded
Wilson subchain of length $n$ (i.e.\ consisting of sites 0 to $n$) and
its equivalent unfolded version, what are the smallest values for the
dimensions $\DNRG$ and $\DDMRG$ for which both approaches describe the
ground state \emph{exactly}, i.e.\ without any truncation?  Answer: On
the one hand, the folded subchain has $n+1$ sites of dimension
$\dNRG$, and hence a total dimension $\dNRG^{n+1}$; to ensure that the
ground state in this space is described exactly, the kept space of the
previous iteration
must 
not involve any truncation, implying $D = \dNRG^n$.  On the other
hand, for the equivalent unfolded subchain, the spin $\downarrow$ and
$\uparrow$ parts each have $n+1$ sites of dimension $\dDMRG$, hence
each have a Hilbert space of total dimension $\dDMRG^{(n+1)}$; to
ensure that this space is described without truncation, its dimension
should equal the maximal dimension of the $\AP$-matrices at sites $0
\mu$, implying $\DDMRG = \dDMRG^{n+1}$. Using $\dDMRG = \sqrt {\dNRG}$
we readily find $\DDMRG = \dDMRG \sqrt \DNRG$, establishing the
proportionality between $\DDMRG$ and $\sqrt{ \DNRG}$ and suggesting
the choice $m= 1$ to achieve an accurate VMPS-representation of the
ground state. Actually, we find numerically that already $m=0$ yields
good qualitative agreement between the VMPS and NRG ground states,
while $m = 1$ yields a \emph{quantitatively} accurate
VMPS-representation of the NRG ground state also for larger chain
lengths, that do involve truncation.  Since such ground states are
built from the kept spaces of previous iterations, this implies that
for $m=1$, all \emph{kept} states in NRG (not only the ground state)
are likewise well represented by VMPS. Indeed, we will find this to be
the case. Moreover, it turns out numerically that with $m=2$, it is
also possible to achieve an accurate VMPS-representation of all kept
\emph{and} discarded NRG-type states, as will be extensively
illustrated below.

For the results reported below, 
we show data only for even iteration number $n$, to avoid
even/odd oscillation effects that are typical and well-understood for
Wilsonian logarithmic discretization, but not of particular interest
here.  We set $\DDMRG = \dDMRG^m \sqrt{\DNRG}$ throughout and specify
the choices made for $m$. All VMPS results shown in this section are
extracted from randomly initialized, fully converged variational
ground states $|E_\ground^N \rangle_\unfolded$ of the form
(\ref{eq:DMRG_MPS}).

\subsection{Ground state energies and overlaps} 
\label{subsec:comparisonenergiesoverlaps}

\begin{figure}[!t]
 \includegraphics[width=1\linewidth]{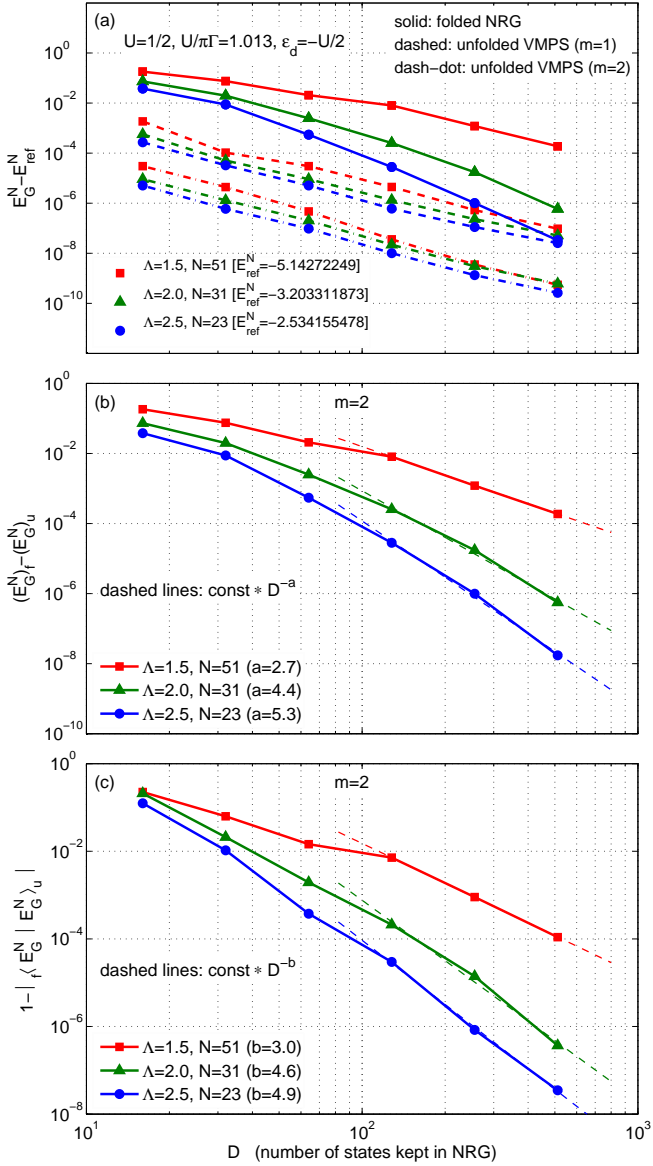} 
 \caption{(Color online) Comparison of NRG and \VMPS\ results for
   (a,b) the ground state energies and (c) the ground state overlaps,
   plotted as a functions of $\DNRG$ with ${\DDMRG}=\dDMRG^m
   \sqrt{\DNRG}$, for three values of $\Lambda$ and, in (a), for two
   values of $m$. In (a) the reference energies
   $E_{\mathrm{ref}}^N$ for each $\Lambda$ were obtained by
   extrapolating the \VMPS\ data points for $m=2$ to
   $\DDMRG\to\infty$, which represents the best estimate of the true
   ground state energy available within the present set of methods.
   The power law fits to the numerical data in (b) and (c), shown as
   dashed lines, were made for the three data points with largest $D$,
   for which the dimensions are large enough to have reliable NRG
   data.}
\label{fig_VMPS(r)}
\end{figure}

Figures~\ref{fig_VMPS(r)}(a) and \ref{fig_VMPS(r)}(b) compare the NRG
and VMPS \emph{ground state energies}, $(E_\ground^N)_\folded$ and
$(E_\ground^N)_\unfolded$, for three values of $\Lambda$ and, in (a),
two values of $m$. They illustrate three points. Firstly, for a given
$\Lambda$ the VMPS ground state energies are smaller than those of
NRG, $(E_\ground^N)_\unfolded < (E_\ground^N)_\folded$, as expected,
since VMPS is a variational method and NRG is not.  Secondly,
Fig.~\ref{fig_VMPS(r)}(a) shows that larger values of $m$ yield lower
VMPS ground state energies, as expected, since their variational space
is larger.  Thirdly, the improvement of VMPS over NRG, as measured by
the energy difference $(E_\ground^N)_\folded -
(E_\ground^N)_\unfolded$ shown in Fig.~\ref{fig_VMPS(r)}(b), becomes
more significant for smaller $\Lambda$, as expected, since the
truncation scheme of NRG relies heavily on energy scale separation,
and hence becomes less efficient for smaller $\Lambda$.

\begin{figure}[tb]
 \centering \includegraphics[width=1\linewidth]{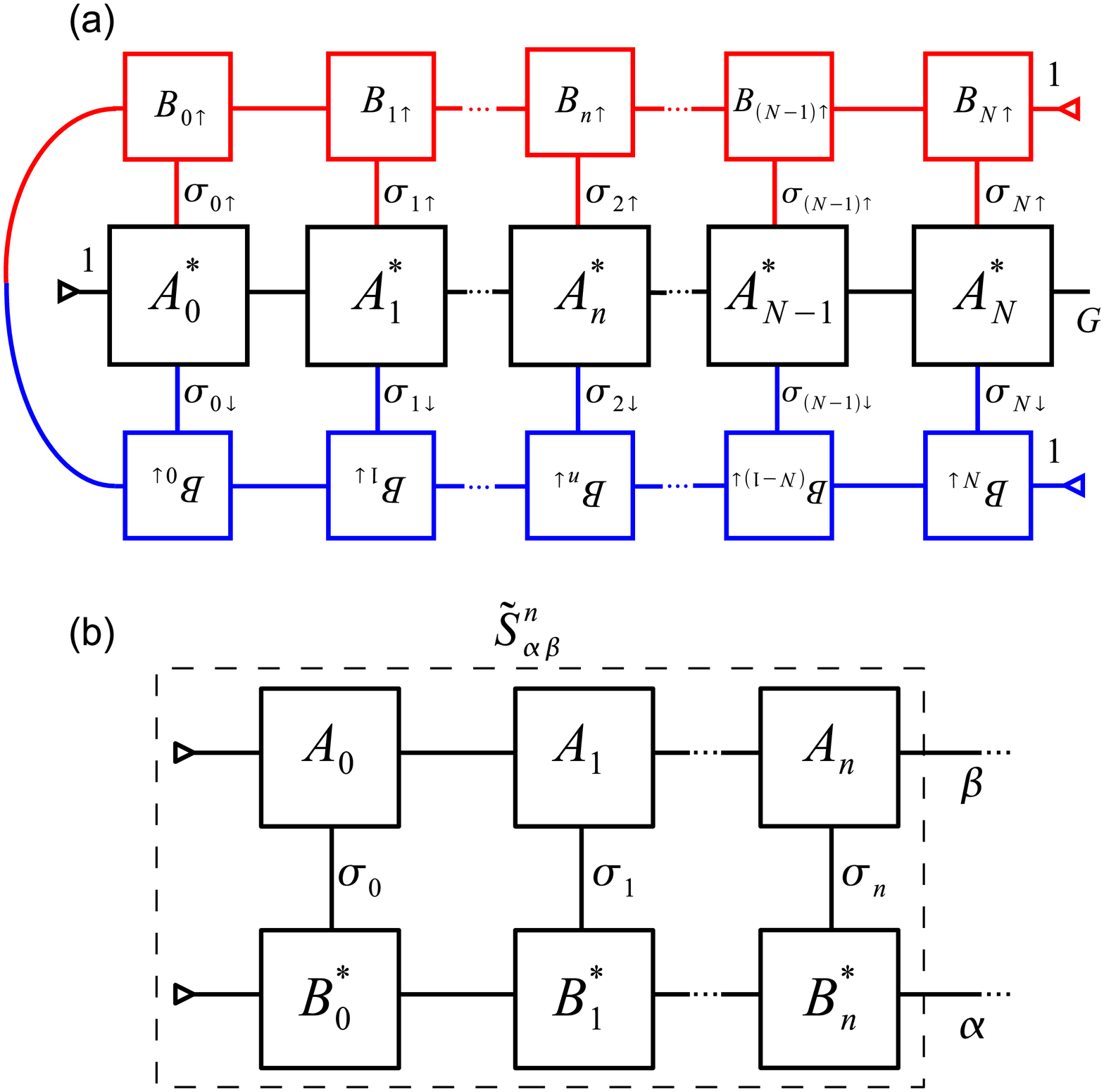} 
 \caption{(Color online) Contraction patterns used to calculate (a)
   the overlap ${}_\folded \langle \EPsi^N_\ground | E^N_\ground
   \rangle_\unfolded $ [\Eq{eq:DMRG-NRGoverlap}] between folded NRG
   and unfolded VMPS ground states, and  (b) the
   overlap matrix $\tilde S_{\alpha \beta}^{n} = {}_\refolded\langle
   \Psi^n_{\alpha}| \EPsi^n_{\beta}\rangle_\folded$
   [\Eq{eq:overlap_matrix_basisstates_contractionpattern}] between
   refolded VMPS basis states and folded NRG eigenstates.  Boxes
   represent $A$ or $\AP$ matrices in the graphical representation of
   Fig.~\ref{fig:matrix-product-structure}, and links connecting them
   represent indices that are being summed over.}
\label{fig_ovpat}
\label{fig:overlap_matrix_basisstates}.
\end{figure}

Figure~\ref{fig_VMPS(r)}(c) compares the overlap between NRG and VMPS
ground states, characterized by the deviation from 1 of the overlap
$|{}_\folded \langle \EPsi^N_\ground | E^N_\ground \rangle_\unfolded|
$.  The latter can be calculated straightforwardly from
\begin{eqnarray}
  \label{eq:DMRG-NRGoverlap}
  {}_\folded \langle \EPsi^N_\ground | E^N_\ground \rangle_\unfolded \!
  & = & 
  \sum_{\lbrace\bsigma^N \rbrace} 
  (A^{[\sigma_N]\dagger} \dots A^{[\sigma_0]\dagger})_{\ground 1} 
\\ 
 & & \times 
  (\AP^{[\sigma_{N\downarrow}]} \dots 
  \AP^{[\sigma_{0\downarrow}]}
  \AP^{[\sigma_{0\uparrow}]} \dots 
  \AP^{[\sigma_{N\uparrow}]})_{11} 
  \nonumber
\end{eqnarray}
where the index contractions associated with the summation over
repeated indices are illustrated in Fig.~\ref{fig_ovpat}(a).
Fig.~\ref{fig_VMPS(r)}(c) shows that the deviation of the overlap from
1 becomes larger the smaller $\Lambda$, again illustrating that then
the NRG truncation scheme becomes less reliable.

\subsection{Comparison of eigenspectra and density of states}
\label{sec:eigenspectraDOS}

\begin{figure*}[tb]
 \includegraphics[width=.85\linewidth]{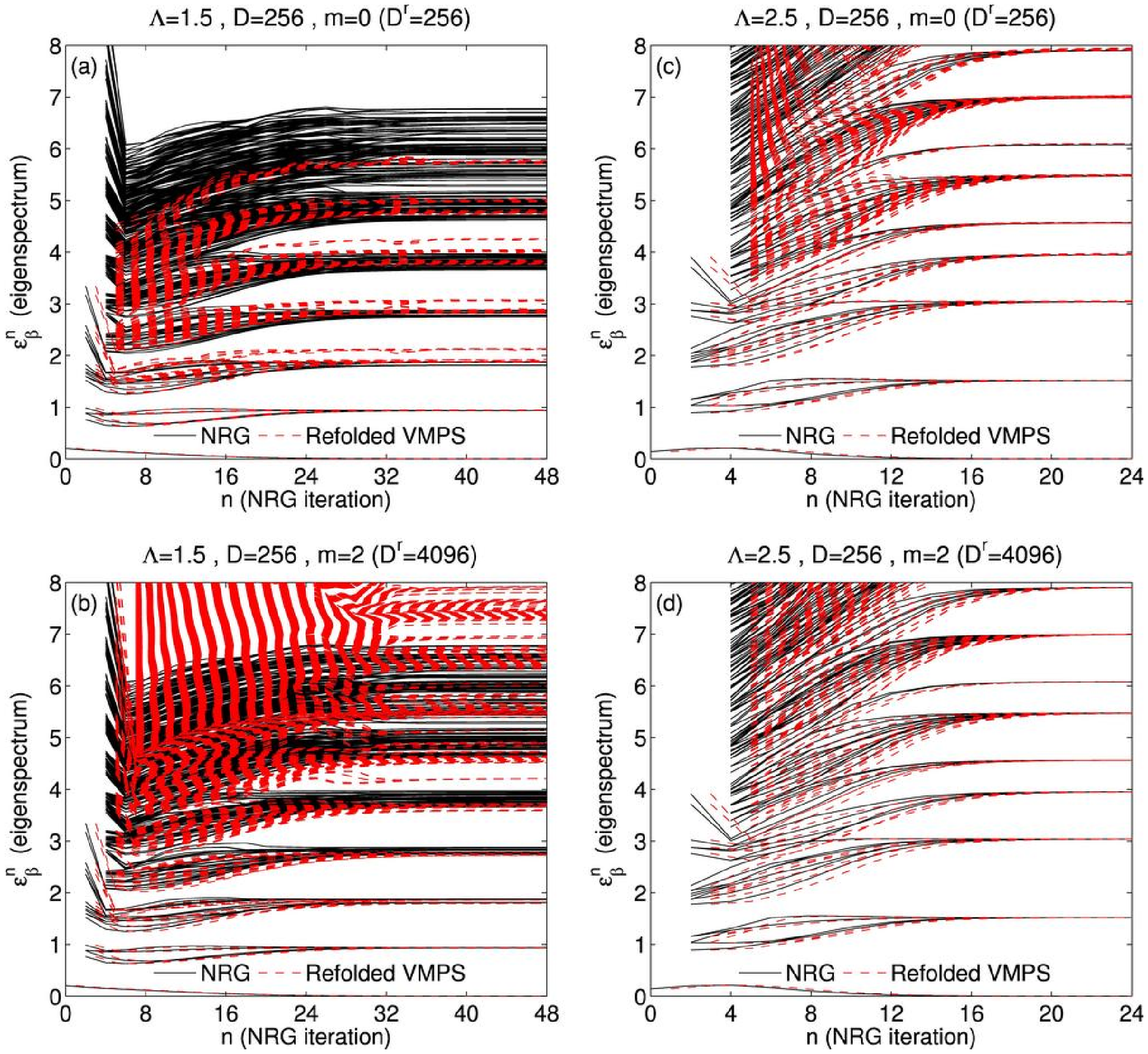} 
 \caption{(Color online) Comparison of energy flow diagrams from NRG
   (dashed red lines) and refolded VMPS data (solid black lines),
   showing the rescaled energies $(\varepsilon_\beta^n)_{\folded,\refolded}$
[\Eq{eq:NRGscaledeigenenergies}] versus $n$, calculated for
   even iteration numbers and four combinations of $m$ (= 0 or 2) and
   $\Lambda$ (= 1.5 or 2.5). The number of NRG states shown (kept
   \emph{and} discarded) is $\DNRG \dNRG$; the number of refolded VMPS
   states shown is $\DNRG^\refolded = \DDMRG^2 = \dNRG^{m} \DNRG$,
   this being the maximal dimension of refolded matrices
   $B^{[\sigma_n]}$. For $m=2$ and $\Lambda = 2.5$, the NRG and DMRG
   flow diagrams agree very well, see (d).  }
\label{fig_flow_diagram}
\end{figure*}

Figure~\ref{fig_flow_diagram} compares the energy \emph{flow diagrams}
obtained from NRG and refolded VMPS data, the latter obtained by
diagonalizing the effective Hamiltonian of \Eq{eq:Heffective}.  It
shows the rescaled energies $({\varepsilon}^n_\beta)_{\folded,\refolded}$
of \Eq{eq:NRGscaledeigenenergies} as functions of $n$,
for four combinations of $m$ and $\Lambda$, and illustrates the same
trends as found in the previous subsection: Firstly, the NRG and VMPS
flow diagrams clearly agree
not only for the ground state but also for
a significant number of excited states. Evidently, the variational
space searched by VMPS is large enough to capture considerable
information about excited states, too, although the VMPS method was
designed to optimize only the ground state.  Moreover, for a given
choice of $\Lambda$, NRG and VMPS eigenenergies coincide for a larger
number of states for $m=2$ than for $m=0$ [compare (b) to (a) and (d)
to (c)], because the variational space is larger. Secondly, for a
given choice of $m$, NRG and VMPS eigenenergies agree better for
$\Lambda=2.5$ than for $\Lambda= 1.5$ [compare (c) to (a) and (d) to
(b)], as expected, because larger $\Lambda$ leads to better energy
scale separation and reduces the inaccuracies inherent in NRG's
Wilsonian truncation scheme.

\label{sec:flow_diag}

\begin{figure}[tb]
 \includegraphics[width=1\linewidth]{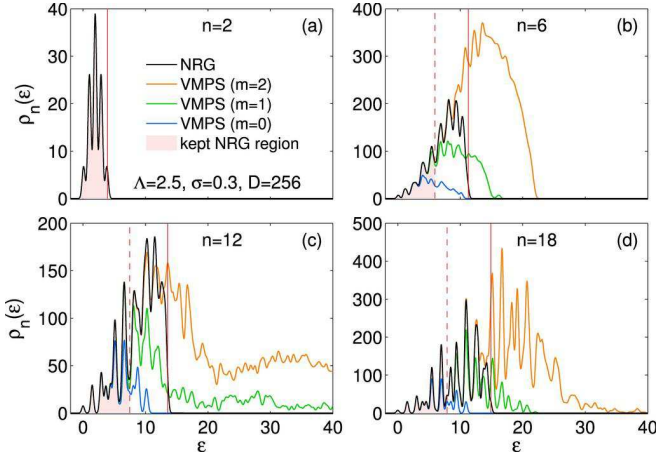} 
 \caption{(Color online) Results for the density of states, $\rho_n
   (\varepsilon)$  [\Eq{eq:DoS}], broadened with a Gaussian broadening function.
 In each panel, the red  vertical dashed and 
solid lines [which coincide in (a)] indicate the energies of the
highest-lying kept and discarded NRG states of that iteration, while
the shaded area indicates the range of kept NRG states.}
\label{fig_DoS_2_5}
\end{figure}

As a complementary way of analysing spectral information 
we also consider the ``density of states'', for a given 
iteration number $n$, 
\begin{eqnarray}
  \DoS(\varepsilon)=\sum_{\alpha=1}^{\DNRG_{\rm max}}
\delta_\sigma (\varepsilon-\varepsilon_{\alpha}^{n}) \; ,
\label{eq:DoS}
\end{eqnarray}
using the rescaled eigenenergies $\varepsilon_\beta^n$ of \Eq{eq:NRGscaledeigenenergies}.
Here $\delta_\sigma (\varepsilon) = e^{-\varepsilon^2/\sigma^2}/(\sigma \sqrt \pi)$ is a
Gaussian peak of width $\sigma$ and unit weight, used to broaden the
discrete spectrum in order to be able to plot it, and the number of
states included in the sum is taken as $\DNRG_{\rm max} = \dNRG \DNRG$
or $\dNRG^m \DNRG$ for NRG or VMPS results, respectively.
Figure~\ref{fig_DoS_2_5} shows such a density of states for several
choices of $m$ and iteration number $n$. It illustrates three points:

Firstly, although for small energies $\DoS (\varepsilon)$ grows rapidly with
$\varepsilon$, as expected for a many-body density of states, it does not
continue to do so for larger $\varepsilon$ (the exact density of states would),
due to the truncation inherent in both NRG and VMPS strategies.  For
NRG, $\DoS (\varepsilon)$ drops to 0 very abruptly, because by construction
Wilsonian truncation is sharp in energy space (at each iteration only
the lowest $\dNRG \DNRG$ eigenstates are calculated). In contrast, for
VMPS $\DoS (\varepsilon)$ decreases more gradually for large $\varepsilon$, because VMPS
truncation for states at site $n$ is based not on their energy, but on
the variationally determined weight of their contribution to the
ground state of the full Wilson chain of length $N$. Evidently, these
weights decrease with increasing $\varepsilon$ less rapidly than assumed by NRG.

Secondly, the agreement of the VMPS curve for $\DoS (\varepsilon)$ with that of
NRG is rather poor for $m=0$ (disagreement sets in already within the
range of kept states of NRG, indicated by the shaded region), better
for $m=1$ (the range of kept states is fully reproduced), and very
good for $m=2$ (disagreement sets in only close to the upper end of
range of discarded states).

Thirdly, for large $n$, $\DoS(\varepsilon)$ becomes increasingly spiky. This
reflects the fact that the spectrum approaches a fixed point with
regularly-spaced eigenenergies, as is evident in the energy flow
diagrams of Fig.~\ref{fig_flow_diagram}.

\subsection{Comparison of energy eigenstates}
\label{sec:eigenstates} 

To compare the energy \emph{eigenstates} produced by NRG and refolded VMPS
for a chain of length $n$, we analyse the overlap
matrix
\begin{eqnarray}
  S_{\alpha \beta}^{n}= {}_\refolded\langle E^n_{\alpha}|
\EPsi^n_{\beta}\rangle_\folded \; . 
\label{eq:overlap_matrix_energystates}
\end{eqnarray}
It can be conveniently calculated from $S^n = U^n \tilde S^n$, where
$U^n_{\alpha \beta} = {}_\refolded \langle E_\alpha^n | \Psi_\beta^n
\rangle_\refolded$ is the matrix that diagonalizes the effective
Hamiltonian matrix $ \H^n_{\alpha \beta} $ of \Eq{eq:Heffective}, and
the matrix
\begin{subequations}
  \label{subeq:Stilde}
\begin{eqnarray}
\label{eq:overlap_matrix_basisstates}
\tilde S_{\alpha \beta}^{n} & = & {}_\refolded\langle \Psi^n_{\alpha}|
\EPsi^n_{\beta}\rangle_\folded \; , 
\\
& = & 
\sum_{\lbrace\bsigma^N \rbrace} 
(\AP^{[\sigma_n]\dagger} \dots \AP^{[\sigma_0]\dagger})_{\alpha 1} 
(A^{[\sigma_0]} \dots A^{[\sigma_n]})_{1\beta} \qqph
\label{eq:overlap_matrix_basisstates_contractionpattern}
\end{eqnarray}
\end{subequations}
characterizes how much weight the NRG eigenstates have in the space
spanned by the refolded VMPS basis states, and vice versa.  The contractions
implicit in \Eq{eq:overlap_matrix_basisstates_contractionpattern} are
illustrated in Fig.~\ref{fig:overlap_matrix_basisstates}(b).

\begin{figure}[tb]
 \includegraphics[width=.95\linewidth]{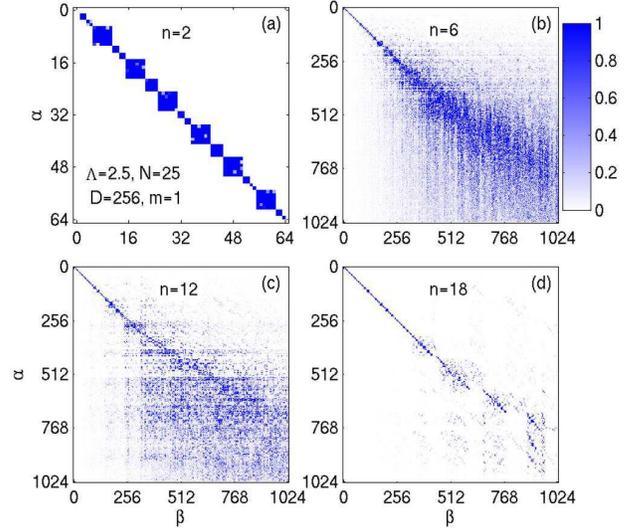} 
  \caption{(Color online) Plot of the overlap matrix $ S_{\alpha
      \beta}^{n}= {}_\refolded\langle E^n_{\alpha}|
    \EPsi^n_{\beta}\rangle_\folded$ [\Eq{eq:overlap_matrix_energystates}]
between refolded VMPS and NRG energy eigenstates, 
with a color scale ranging between 0 and 1.
In (a), with $n=2$, no truncation occurs at all, 
and both state labels $\alpha$ and $\beta$ 
run from 1 to $\dNRG^{n+1} = 64$. In (b) to (d),
truncation does occur: For the folded NRG eigenstates 
$|\EPsi^n_{\beta}\rangle_\folded$, the label $\alpha$ runs from
1 to $\DNRG \dNRG = 1024$, i.e.\ it includes all kept and discarded NRG 
states, while for the refolded VMPS eigenstates 
$|E^n_\beta \rangle_\refolded$, the label $\beta$ 
runs from 1 to $\DNRG^\refolded_n = 
\DDMRG^2 = 1024$ [\Eq{eq:DNRG-refolded}]. }
\label{fig_plotted_ov_mat}
\end{figure}

Figure~\ref{fig_plotted_ov_mat} shows the overlap matrix $S^n_{\alpha
  \beta}$ on a color scale ranging from 0 to 1, for $m = 1$ and
several values of $n$. For the region of low excitation energies
(about the first hundred or so states) its structure is evidently close to
block-diagonal, indicating that both sets of states from which it is
built are reasonably good energy eigenstates.  Had both sets been
perfect energy eigenstates, as would be the case for $\DDMRG$ and
$\DNRG$ large enough to avoid all truncation, the blocks would be
completely sharp, with sizes determined by the degeneracies of the
corresponding energies. Sharp blocks are indeed observed for $n=2 $ 
[Fig.~\ref{fig_plotted_ov_mat}(a)], because no truncation has occurred yet. 
The ``fuzziness'' shown by the blocks in
Fig.~\ref{fig_plotted_ov_mat}(b) to \ref{fig_plotted_ov_mat}(d) 
for larger $n$ implies that truncation
is beginning to make itself felt, causing NRG and VMPS to increasingly
disagree on how to construct the eigenstates corresponding to a given
range of eigenenergies. Note that the fuzziness becomes 
markedly more pronounced for $\alpha, \beta  > 256$.
The reason  is that whenever $S^n_{\alpha\beta}$ is
nonzero for $\beta>\DNRG$, the associated VMPS states have weight
among the discarded states of NRG, implying that NRG discards some
states relevant for building the VMPS ground state.
Thus,
$S^n_{\alpha\beta}$ quite literally measures to what extent the
truncation criteria of NRG and VMPS are compatible.  
Near the end of the chain, for $n = 18$
[Fig.~\ref{fig_plotted_ov_mat}(d)], the off-diagonal spread
is significantly reduced compared to the middle of the chain $(n=6,12)$
[Fig.~\ref{fig_plotted_ov_mat}(b,c)], for two reasons.
Firstly, the dimensions of the refolded $B$-matrices 
become small for $n$ near $N$, see
\Eq{eq:DNRG-refolded}, so that the amount
of truncation is much less severe near the end of 
the chain than in its middle. Secondly, the eigenspectra
have converged to their fixed point values, so that the
number of different eigenenergies in a given energy
interval is reduced, thus reducing the fuzziness in 
Fig.~\ref{fig_plotted_ov_mat}(d).

\begin{figure}[tb]
 \includegraphics[width=1\linewidth]{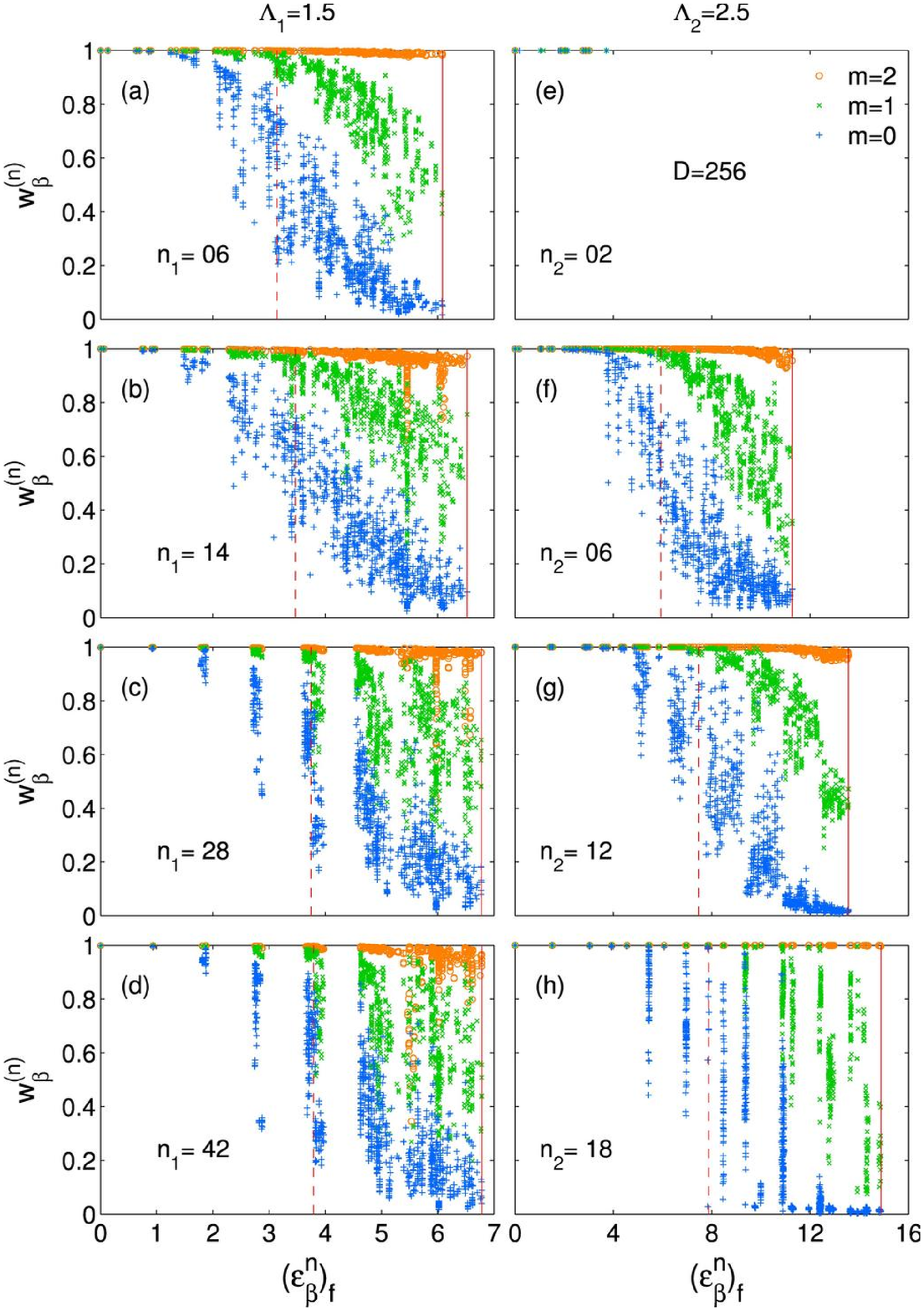} 
  \caption{(Color online) For several NRG iteration numbers $n$ and
    two values of $\Lambda$ (different panels), this figure shows the
    weights $w_{\beta}^{(n)}$ [\Eq{eq:wiD}] with which NRG states
    $|\EPsi_\beta^n \rangle_\folded $ with rescaled NRG eigenenergies
    $(\varepsilon_\beta^{n})_\folded$ [\Eq{eq:NRGscaledeigenenergies}]
    are found to lie in the VMPS-subspace of dimension $\DDMRG =
    \dDMRG^m \sqrt \DNRG$, with $m = 0$, 1 or 2 (indicated by $+$,
    $\times$ or $\circ$, respectively).  In each panel, the red
    vertical dashed and solid lines indicate the energies of the
    highest-lying kept and discarded NRG states of that iteration. For
    $n=3$, both of these lines are missing, since truncation has not yet
    set in. The choices for $n$ in the left and right panels of each
    row are related by
$\Lambda_1^{-n_1/2} = \Lambda_2^{-n_2/2}$, to 
ensure that both panels show a comparable energy scale. 
 }
\label{fig_Wi_Ei2}
\end{figure}


Next consider the total 
weight which a given  NRG-state $|\EPsi^n_\beta\rangle_\folded$ 
has within the refolded VMPS-subspace for that $n$, 
\begin{eqnarray}
w_{\beta}^{(n)}=
\sum_{\alpha=1}^{\DNRG_n^\refolded}
|S_{\alpha\beta}|^{2}=\sum_{\alpha=1}^{\DNRG_n^\refolded}
|{\tilde S}_{\alpha\beta}|^{2}\;.
\label{eq:wiD}
\end{eqnarray}
It satisfies $0\le w_{\beta}^{(n)}\le1$. Weights less than 1 imply
that the VMPS-subspace is too small to adequately represent the
corresponding NRG state. The second equality in \Eq{eq:wiD}, which
follows from the unitarity of $U$, is useful since it implies that
these weights can also be calculated directly from the refolded states
$|\Psi_\beta^n \rangle_\refolded$ [\Eq{eq:refolded_DMRG_MPS_basis}], without
the need for diagonalizing the large ($\DDMRG^2 \times
\DDMRG^2$-dimensional) effective refolded Hamiltonian
 $\H^n_\refolded$
[\Eq{eq:Heffective}].

Figure~\ref{fig_Wi_Ei2} shows such weights $w^{(n)}_\beta$ for various
choices of $n$, $\Lambda$ and $m$.  Their dependence on $m$ reinforces
the conclusions of the previous subsection: For $m=0$ (blue $+$
symbols), the weights are equal to 1 for the lowest state
of each iteration, but less than 1 for many of the
kept states. This shows that the VMPS subspace is large enough
to accurately
represents the NRG ground state, but  significantly too small to
accurately represent all kept states.  For $m=1$ (green $\times$
symbols), the weights are close to 1 only for the kept states, while
smoothly decreasing towards 0 for higher-lying discarded states.
Finally, for $m=2$ (orange $\circ$ symbols), the weights of both kept
and discarded NRG states are all close to 1, implying that the VMPS
subspace is large enough to accurately represent essentially
\emph{all} states kept track of by NRG.  Note that for $m=0$ and 1,
the decrease of the weights $w^{(n)}_\beta$ with increasing energy
occurs in a smooth and gradual fashion, illustrating yet again the
smooth nature of VMPS truncation when viewed in energy space.  When a
smaller value of $\Lambda$ is used [compare panels (a-d) to (e-h)] the
weights of the higher-lying states of a given iteration tend to spread
out over a larger range of values, since NRG has a weaker energy scale
separation for smaller $\Lambda$.  Finally, the increasing spikyness
of the eigenspectrum with increasing $n$, see
Fig.~\ref{fig_Wi_Ei2}(d,h), is due to the approach to a fixed point
spectrum with regularly-spaced eigenenergies, as mentioned above.

\begin{figure}[tb]
 \includegraphics[width=1\linewidth]{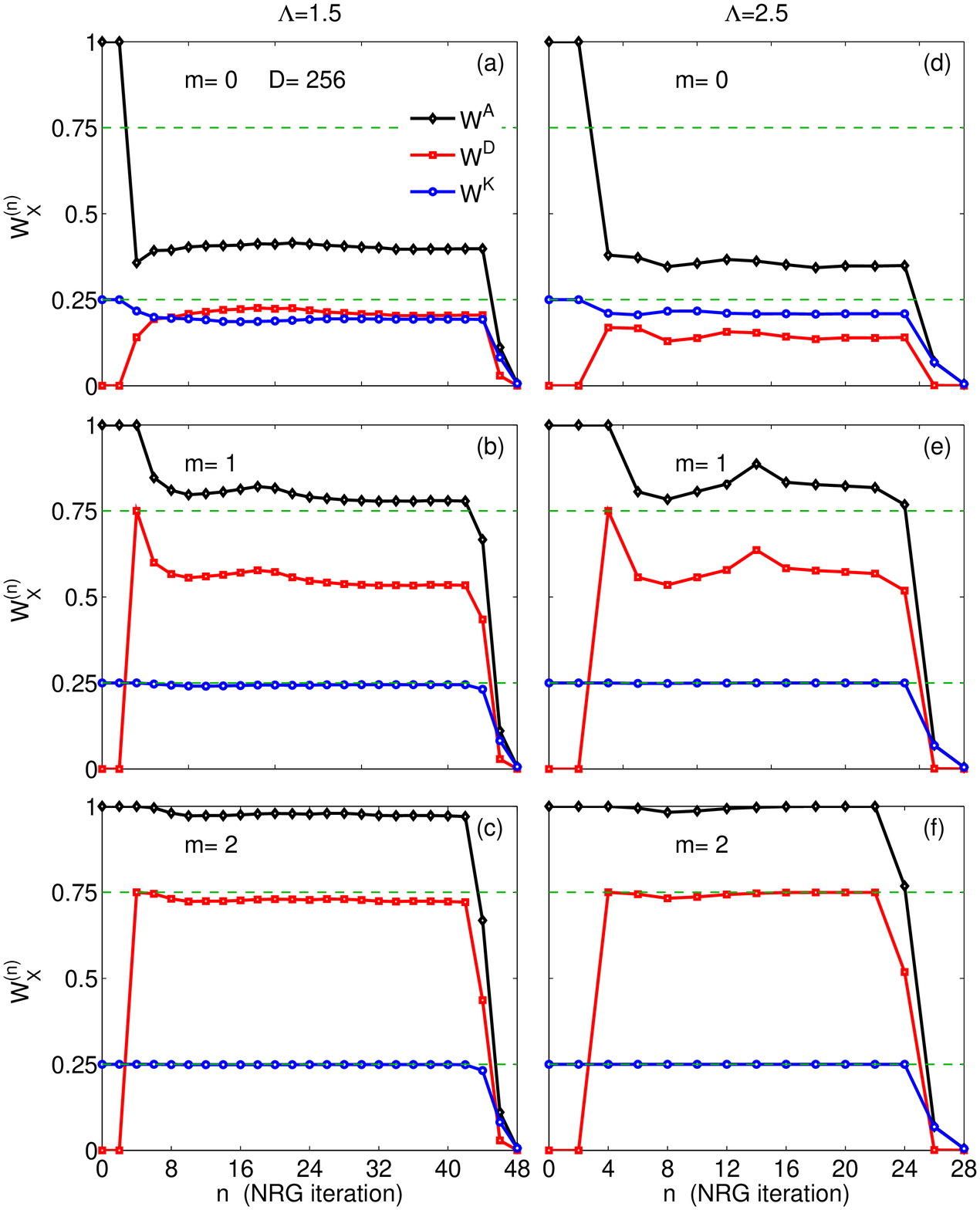}

 \caption{(Color online) Integrated weights $W_{\XX}^{(n)}$ (see
   \Eq{eq:int_w}) for two different $\Lambda$ and three values of $m$.
   Dashed lines depict the maximum possible values of the
   kept and discarded weights, ${1\over 4}$ and ${3 \over 4}$ for
   $W_{\Kept}^{n}$ and $W_{\Discarded}^{n}$, respectively.}
\label{fig_int_w}
\end{figure}

The results just discussed may be represented more compactly 
by considering, for a given iteration $n$, the 
integrated weights obtained by
summing up the weights of \emph{all} NRG states of type $\XX$,
\begin{eqnarray}
  W_{\XX}^{(n)}={1 \over \dNRG \DNRG} 
\sum_{\beta \in \XX}    w_{\beta}^{(n)}
\; ,
\label{eq:int_w}
\end{eqnarray}
where $\XX= \Kept, \Discarded, \Total$ stands for \emph{kept},
\emph{discarded} or \emph{all}, respectively. All three types of
integrated weights are normalized to the total number $\dNRG \DNRG$ of
all NRG states calculated at a given iteration (with ${\dNRG}=4$
here), and reach their maximal values (${1 \over 4}$, ${3 \over 4}$
and $1$, respectively) when all the individual weights for that
iteration equal 1.  Figure~\ref{fig_int_w} shows such integrated weights
for several values of $m$ and $\Lambda$. Upon increasing $m$ from 0 to
2, the integrated weights tend toward their maximal values, doing so
more rapidly for larger $\Lambda$. For $m=2$, they essentially
saturate their maximal values, indicating yet again that the
VMPS variational space is now large enough to fully retain
all information kept track of by NRG.

To summarize the result of this section: The VMPS approach reproduces
NRG ground state properties much more cheaply, requiring only $\DDMRG
= \sqrt \DNRG$ for qualitative agreement, and $\DDMRG = \dDMRG \sqrt
\DNRG$ for quantitative agreement.  Moreover, it can also reproduce
all kept and discarded NRG eigenstates if $\DDMRG = \dDMRG^2 \sqrt
\DNRG$ is used.  However, to obtain excited energy eigenstates, we
have to refold, requiring the diagonalization of matrices of dimension
$\DDMRG^2 \times \DDMRG^2$.  The numerical cost of doing so is
comparable to that of NRG.

The fact that VMPS gives access to the same information on eigenstates
and eigenvalues as NRG has a very significant and reassuring
consequence: \emph{all} physical properties of the model that can be
calculated by NRG can also be calculated by VMPS, in combination with
refolding.

\section{Cloning and variational improvement of NRG ground state
\label{sec:cloning+optimization}}

Viewed in  MPS language, the NRG method constructs the ground state
in a single sweep along the chain: each $A$ is calculated
only once, without allowing for possible feedback of information from
$A$'s describing lower energies to those of higher energies calculated
earlier. Thus, the resulting NRG ground state $|E_G^N \rangle_\folded $,
to be denoted simply by $| \GG \rangle_\folded $ below,
is not optimal in a variational sense. In this section
we investigate to what extent the ground 
state energy can be lowered further by performing 
variational energy optimization sweeps on $| \GG \rangle_\folded $ 
that serve to
account for feedback of information from low
to high energy scales. This feedback  turns out to be small
in practice, as will be seen below, but it is not strictly
zero and its importance increases as the logarithmic discretization
is refined by taking $\Lambda\to1$.

\subsection{Mapping folded to unfolded states by cloning}
\label{sec:cloning} 

Our first step is to rewrite a given NRG ground state $|\GG
\rangle_\folded $ in a form amenable to subsequent energy minimization
sweeps. To this end, we use a variational \emph{cloning} procedure
(subscript c),
\begin{eqnarray}
  \label{eq:cloningmap}
  |\GG \rangle_\folded  \stackrel{{\rm cloning}}{\longrightarrow}
|\GG \rangle_\cloned \in \{ |\Psi^N \rangle_\unfolded  \} \; , 
\end{eqnarray}
which maps $|\GG \rangle_\folded $ of the form of \Eq{eq:NRG_MPS}
[Fig.~\ref{fig:matrix-product-structure}(a)] onto an unfolded state
$|\GG \rangle_\cloned $ of the form $|\Psi^N \rangle_\unfolded $ of
\Eq{eq:DMRG_MPS} [Fig.~\ref{fig:matrix-product-structure}(c)].  Since
their matrix product structures differ, this mapping will, for general
values of $\DNRG$ and $\DDMRG$, not be exact, though its accuracy
should improve systematically with increasing $\DDMRG$ and hence
increasing dimensions of the variational space.  To be explicit, we
seek the best possible approximation to $|\GG \rangle_\folded$ in the
space of all unfolded states of the form ({\ref{eq:DMRG_MPS}), by
  solving the minimization problem
\begin{eqnarray}
\label{eq:distance}
  \min_{|\GG \rangle_\cloned \in \{ |\Psi^N \rangle_\unfolded \} } 
\left[ \parallel |\GG \rangle_\folded - |\GG \rangle_\cloned \! \parallel^{2}
+ \lambda ( 
\parallel \! | \GG \rangle_\cloned \! \parallel^2 -1 )\right] 
\;,
\end{eqnarray}
which minimizes the ``distance'' between $| \GG \rangle_\cloned$ and
$| \GG \rangle_\folded$ under the constraint, implemented using a
Lagrange multiplier $\lambda$, that the norm ${}_\cloned \langle \GG |
\GG \rangle_\cloned = 1 $ remains constant. Varying \Eq{eq:distance}
with respect to the matrices  defining $| \GG \rangle_\cloned$ 
leads to a set of
equations, one for each $k \mu$, of the form
\begin{eqnarray}
\frac{\partial}{\partial B^{[\sigma_{k\mu}]}}
\Bigl[
(1+\lambda) \: 
{}_\cloned \langle \GG |  \GG \rangle_\cloned 
- 2 {\rm Re} \bigl( {}_\folded   \langle \GG | \GG  \rangle_\cloned 
\bigr)  \Bigr]
=0  , 
\label{eq:optim_problem}
\end{eqnarray}
which determine the $B$-matrices of the desired ``cloned'' state $| \GG
\rangle_\cloned$.  These equations can be solved in a fashion entirely
analogous to energy optimization: Pick a particular site of the
unfolded chain, say $k\mu$, and solve the corresponding 
\Eq{eq:optim_problem} for the matrix $B^{[\sigma_{k\mu}]}$ while
regarding the matrices of all other sites as fixed. Then move on to
the neighboring site and in this fashion sweep back and forth through
the chain until convergence is achieved. Appendix~\ref{app:cloning}
describes some details of this procedure.

\label{sec:cloning_results}

A figure of merit for the success of cloning
is the deviation of the 
overlap
$|{}_\cloned \langle \GG | \GG \rangle_\folded| $ from 1.  This
deviation decreases monotonically with successive cloning
sweeps and converges to a small but finite ($\DDMRG$-dependent) value 
when the cloning process converges, as
illustrated in the inset of Fig.~\ref{fig:cloning}.  The main part of
Fig.~\ref{fig:cloning} shows that when the number $\DDMRG$ of VMPS
states is increased, the converged value of the overlap deviation
approaches 0 as a power law in $\DDMRG$ (red circles).  It also shows
that the corresponding VMPS truncation error $\tau (\DDMRG)$ incurred
during cloning (blue squares), calculated according to
\Eq{eq:truncationerror}, likewise decreases in power-law fashion with
$\DDMRG$.  All in all, Fig.~\ref{fig:cloning} confirms that cloning
works very well if $\DDMRG$ is sufficiently large.

\begin{figure}[tb]
 \centering \includegraphics[width=1\linewidth]{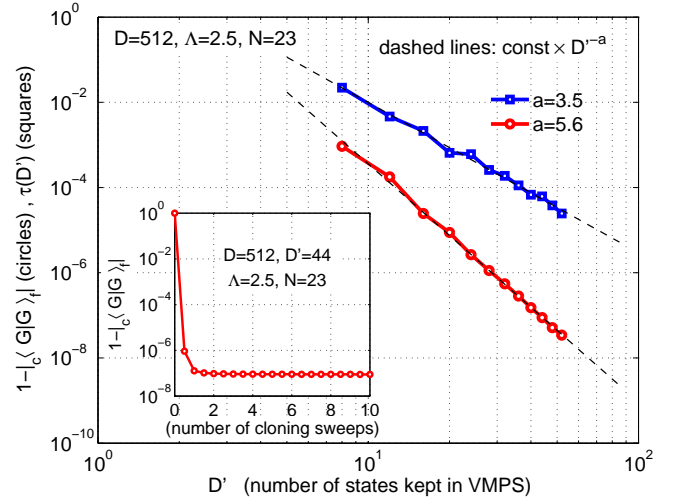} 
 \caption{(Color online) The deviation of the overlap $|{}_\cloned
   \langle \GG | \GG \rangle_\folded |$ from 1 (red circles) and the
   cloning truncation error $\tau (\DDMRG)$ (blue squares), as
   functions of the number $\DDMRG$ of kept states in the cloning
   procedure. Both approach 0 in power-law fashion, as indicated by
   the dashed line fits.  The inset shows how the overlap deviation
   from 1 decreases and converges to a small but finite constant in
   the course of sequential cloning sweeps.  }
\label{fig:cloning}
\end{figure}

\subsection{Variational energy minimization of $|\GG\rangle_\cloned$}
\label{sec:sw_on_NRG}

\begin{figure}[tb]
 \includegraphics[width=0.95\linewidth]{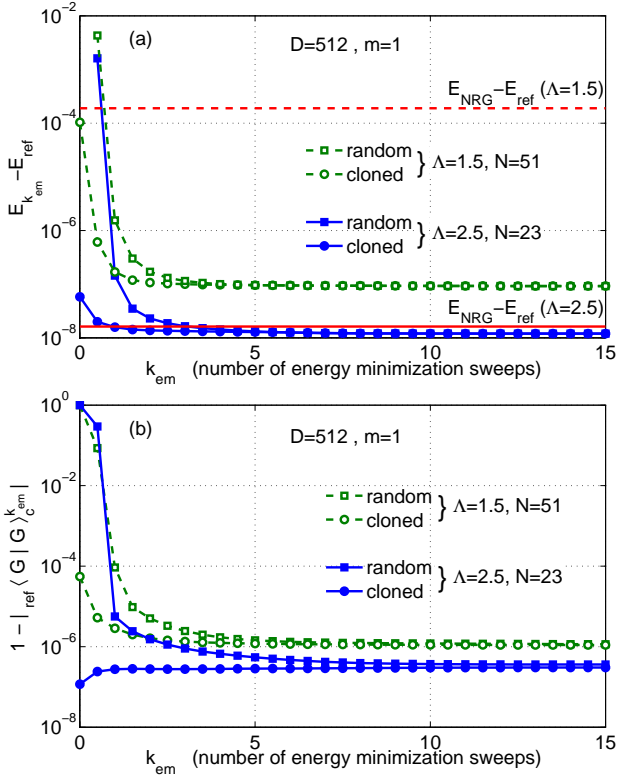} 
 \caption{(Color online) Comparison of (a) energies and (b) wave
   function overlaps for random initialization (squares) vs.\
   NRG-cloned initialization (circles), as functions of the number
   $\kem$ of variational energy minimization sweeps.  Results are
   shown for $\Lambda = 1.5$ (green, open symbols, dashed lines) and
   $\Lambda = 2.5$, (blue, filled symbols, solid lines).  The energies
   in (a) and overlaps in (b) are calculated with respect to a
   reference ground state $|\GG\rangle_\reference$ with $\DDMRG = 64$,
   obtained by performing 50 energy minimization sweeps starting from
   random initialization. The red horizontal straight lines in (a)
   (dashed or solid for $\Lambda = 1.5$ or 2.5, respectively), show
   the energy difference $E_\NRG - E_\reference$, where $E_\NRG$ is
   the energy of the NRG ground state $|\GG \rangle_\folded$ used as
   input into cloning.  The fact that $E_\NRG $ does not completely
   coincide with the energy $E_\cloned = E_{\kem=0}$ of the cloned
   state (horizontal straight lines do not meet circles at $\kem =0$)
   is due to the fact that the deviation of the  overlap
   $|{}_\cloned \langle \GG | \GG \rangle_\folded| $ from 1 is not
   strictly equal to 0 (see Fig.~\ref{fig:cloning}).  }
\label{fig_swoNRG}
\end{figure}
\label{sec:convergence-energyoptimization}

Having used cloning to find the optimal unfolded representation
$|\GG\rangle_\cloned$ of the NRG ground state $|\GG\rangle_\folded$, we
now variationally minimize its
energy by sweeping. We thereby obtain a sequence
of states $|\GG\rangle_\cloned^\kem$ of ever lower energy, 
$E_\kem$, where the index
$\kem = 0, 1, 2, \dots $ gives the number of energy minimization
sweeps that have been 
performed. The procedure is precisely analogous to
that described in Section~\ref{sec:MPSAnsatz},
the only difference being that the random initial
state used there is here replaced by the cloned state
$|\GG\rangle_\cloned^0 = |\GG\rangle_\cloned$ as initial state.

Figure~\ref{fig_swoNRG}(a) shows the evolution of the ground state
energy $E_\kem$ as function of the number $\kem$ of energy
minimization sweeps, for both random (squares) and cloned (circles)
initial states. $E_\kem$ is displayed with respect to the energy
$E_\reference$ of a reference state $|\GG \rangle_\reference$, defined
in the figure caption, which represents our best approximation to the
true ground state.  Figure~\ref{fig_swoNRG}(b) shows how $1-
|{}_\reference \langle \GG | \GG \rangle_\cloned| $ decreases as
sweeping procedes, converging to a small but
finite value.  For a given value of $\Lambda$ (1.5, shown in
green, open symbols connected by dashed lines, or 2.5, shown in blue,
filled symbols connected by solid lines), the energies for random and
cloned initialization shown in Fig.~\ref{fig_swoNRG}(a) converge to
the same value within just a few sweeps.  However, the convergence is
quicker for the cloned (circles) than the random (squares) input
state, since the former represents an already rather good initial
approximation (namely that of NRG) to the true ground state, whereas
the latter is simply a random state.  Nevertheless, the circles show
strikingly that the NRG ground energy is \emph{not} optimal, in that
the energy can be lowered still further by sweeping.  Moreover, this
improvement is more significant for small than large $\Lambda$ (for
circled data points, compare dashed green to solid blue lines for
$\Lambda=1.5$ or 2.5, respectively). The reason is that the NRG
truncation scheme becomes less accurate the smaller $\Lambda$ is,
implying that the NRG result can be improved more significantly by
further sweeping.  This is again a reminder that the systematic error
of NRG increases as $\Lambda$ approaches 1, as already observed in
Fig.~\ref{fig_VMPS(r)}.

\section{Conclusions}
\label{sec:conclusions}

In this paper we presented a systematic comparison between NRG and
DMRG, which we mainly referred to as VMPS, for the single-impurity
Anderson model within the framework of matrix product states. We first
reformulated both NRG and DMRG in the language of MPS, using a folded
Wilson chain for NRG and an unfolded one for DMRG.  Then we
quantitatively compared the results of NRG and the VMPS approach for
energy eigenvalues and eigenstates and explicitly analysed the
difference in their truncation criteria, which are sharp or smooth in
energy space, respectively.

The most important conclusion of our study is this: For the purpose of
obtaining the ground state of this model, the VMPS approach applied to
the unfolded Wilson chain yields a very significant increase in
numerical efficiency compared to NRG (${\DDMRG}=\dDMRG \sqrt{\DNRG}$),
essentially without loss of relevant information. The physical reason
is that the spin-down and -up chains are only weakly entangled for
this model, so that the NRG matrices $A^{[\sigma_n]}$ of dimension
$\DNRG$ that describe site $n$ of the Wilson chain, can, in effect, be
factorized as a direct product $B^{[\sigma_{n\downarrow}]} \otimes
B^{[\sigma_{n\uparrow}]}$ of two matrices, each having dimension
$\dDMRG \sqrt \DNRG $.  It should be emphasized, though, that this
property relies on the physics of the model, namely the weak
entanglement of the spin down and up chains. To what extent this
property survives for other impurity models should be a subject for
further research, the two-channel Kondo model being a particularly
interesting candidate in this respect. 

Nevertheless, the 
possibility of using unfolded Wilson chains to reduce numerical
costs for ground state calculations is very attractive 
for possible applications of the
VMPS method to more complicated models involving more than
one conduction band.\cite{Holzner} For example, the conductance through
a quantum dot coupled to two leads can under 
certain conditions (linear response, zero temperature,
Fermi liquid behavior, etc.) be expressed in terms a set of phase shifts
that are uniquely determined by the ground state occupation of the dot
energy levels.\cite{PustilnikGlazmanJPhysCM04} 
Thus, in such situations reliable knowledge
of the ground state is sufficient to calculate transport properties.

Going beyond ground state properties, we showed that the entire
excited state eigenspectrum of both kept and discarded NRG states can
be recovered within the VMPS approach with at least the same accuracy
as NRG, by using ${\DDMRG}={\dDMRG}^{2}\times\sqrt{\DNRG}$ and
refolding.  However, the latter step requires a subsequent additional
diagonalization of matrices of dimensions $\DDMRG^2$,
giving rise to a significant increase in numerical resources compared
to the case that only ground state information is required.  A
quantitative comparison between NRG and VMPS for the eigenspectrum's
energies and eigenstates showed better agreement for $\Lambda = 2.5$
than 1.5, due to the fact that the NRG truncation scheme becomes
increasingly less accurate the closer $\Lambda$ approaches 1.

Finally, we used a cloning procedure to recast a given folded NRG
ground state into an unfolded form, and showed that its energy could
be lowered further by subsequent energy minimization sweeps.  As
expected, we found that sweeping improves the relative accuracy with
which the ground state energy can be determined, the more so the
smaller the value of $\Lambda$.  For example, for $\Lambda = 1.5$ the
accuracy changed from ${\cal O} (10^{-4})$ before sweeping to ${\cal
  O} (10^{-7})$ thereafter [see Fig.~\ref{fig_swoNRG}(a)].  The fact
that such a further variational improvement of the NRG ground state is
possible, however, is of significance mainly as a matter of principle,
not of practice: for the numerous situations where NRG works well (in
particular, for $\Lambda$ not too close to 1), we expect that such
further variational improvement of the NRG ground state will not
noticeably affect any physical observables.

Let us conclude with some comments about the pros and cons of NRG and
VMPS. For quantum impurity models with a comparatively low degree of
complexity, such as the single-lead Anderson and Kondo models, NRG
works exceedingly well and for practical purposes nothing is to be
gained from switching to the VMPS approach. The latter is a
potentially attractive alternative to NRG only for  two
types of situations, namely (i) more complex quantum
impurity models, and (ii) non-logarithmic discretization 
of the leads. We briefly discuss these in turn.

(i) For complex quantum impurity models, in particular ones involving
several leads, VMPS achieves a very significant reduction in memory
cost, relative to NRG, for describing ground state properties via
unfolding the Wilson chain.  There are several caveats, though.
Firstly, this reduction in memory cost applies only when \emph{only}
ground state properties are of interest. To obtain excited state
eigenspectra, the memory costs of NRG and VMPS are comparable.
Secondly, unfolding is expected to work well only for models for which
the subchains that are being unfolded are only weakly entangled, which
will not be the case for all impurity models. For example, the
two-channel model might be an example where unfolding works less
well. In general, one needs to check the extent to which degrees of
freedom on different subchains are entangled with each other, by
calculating the mutual information of two sites on different
subchains. If this does not decrease rather rapidly with their
separation from the impurity site, then unfolding will be a poor
strategy. Appealingly, though, such a check can be done purely using
NRG data, as illustrated in Section~\ref{sec:mutualinformation}.
Thirdly, the fact that VMPS relies on variationally optimizing the
ground state might cause convergence problems for models which have
degenerate ground states.  Conceivably this problem can be reduced by
systematically exploiting all relevant symmetries of the Hamiltonian,
including non-Abelian
symmetries,\cite{TothZarandArXiV08,McCouloghArXiV08}.  However, if
states in the local state space of a folded Wilson chain are related
by a non-Abelian symmetry, then this symmetry would not be manifest
in the unfolded representation. Thus, the two possible strategies for
achieving significant memory reduction, namely unfolding and
exploitation of symmetries might not always be mutually compatible;
which one is more favorable
will depend on the details of the model, and is an interesting
subject for further study.

(ii) The VMPS approach offers clear advantages over NRG in situations
where Wilson's logarithmic discretization of the conduction band
cannot be applied. In the present paper, we found clear indications
for this fact in the observation that the improvement of VMPS relative
to NRG becomes more significant as $\Lambda$ is chosen closer to 1.
More importantly, VMPS offers the possibility, inaccessible to NRG, to
improve the frequency resolution of spectral functions at high
frequencies, by using a flexible (non-logarithmic) discretization
scheme which reduces the level spacing of effective lead states in the
energy regimes where higher frequency resolution is desired.  For such
a discretization scheme Wilsonian energy scale separation is lost and
NRG truncation cannot be applied. However, the ground state can still
be found variationally, and spectral functions can be computed using
projection operator techniques. In this fashion, it has recently been
possible to calculate the spectral function for the Anderson model at
large magnetic fields, $B > T_{\rm K}$, and to resolve the split Kondo
resonance with sufficient accuracy to reproduce the widths expected
from perturbation theory in this regime. These developments, though,
go beyond the scope of the present paper and will be published
separately.\cite{VMPS,Muender}

\begin{acknowledgments}
  We gratefully acknowledge fruitful discussions Frithjof Anders,
  Theresa Hecht, Andreas Holzner, Ulrich Schollwöck, Frank Verstraete
  and Gergely Zar\'{a}nd, and thank Frithjof Anders for constructive
  comments on the manuscript.  This work was supported by the
  Spintronics RTN and the DFG (SFB 631, SFB-TR12, De-730/3-2).
  Financial support of the German Excellence Initiative via the
  Nanosystems Initiative Munich (NIM) is gratefully acknowledged.
\end{acknowledgments}

\appendix

\section{Technical Details}
\label{app:A}

In this appendix, we collect some technical details on various
manipulations involving matrix product states.  

\subsection{Orthonormalization of $\AP$-matrices of unfolded Wilson
  chain}
\label{sec:orthonormalization}

To keep the notation simple, in this subsection we shall imagine the
sites of the unfolded Wilson chain  to be stretched along a
line running from left to right,  enumerated by an index $k$
running from 1 for site $N\!\! \!\downarrow$ to $K =2(N+1)$ for site
$N\!\! \! \uparrow$. 
Correspondingly, matrix product states will generically be
written as $|\Psi \rangle = \sum_{\{ \bsigma^K \} } |\bsigma^K \rangle
(\prod_{k=1}^K \AP^{[\sigma_k]})$, with  matrix elements 
$\AP^{[\sigma_k]}_{\nu\eta}$.

It is convenient to ensure that every $\AP$-matrix in a matrix product
state satisfies one of the following two 
orthonormality conditions:
\begin{subequations}
\label{subeq:orthonormality}
\begin{eqnarray}
  \sum_{\sigma_{k}}\AP^{[\sigma_{k}]\dagger}\AP^{[\sigma_{k}]}
& = &
\mathds{1} \; ,  
\label{eq:orthonormality-left-to-right}
\\
  \sum_{\sigma_{k}}\AP^{[\sigma_{k}]}\AP^{[\sigma_{k}]\dagger}
& = &
\mathds{1} \; .
\label{eq:orthonormality-right-to-left}
\end{eqnarray}
\end{subequations}
In particular, if \emph{all} $\AP$-matrices satisfy either the first or the
second of these conditions, the corresponding matrix product state is
automatically normalized:
\begin{equation}
\langle \Psi | \Psi \rangle = \sum_{\{ \bsigma^K \} } 
(\AP_{1 \nu'}^{[\sigma_K] \dagger} \dots \AP_{\eta' 1}^{[\sigma_1]\dagger})
(\AP_{1 \eta}^{[\sigma_1]} \dots \AP_{\nu 1}^{[\sigma_K]}) = 1 \; . 
\label{eq:normalizationsatisfied}
\end{equation}
This follows by iteratively applying \Eq{subeq:orthonormality}.  To
start the iteration, note that for matrices at the beginning or end of
the chain, where one of the matrix indices is a dummy index with only
a single value, \Eqs{eq:orthonormality-left-to-right} or
(\ref{eq:orthonormality-right-to-left}) imply $\sum_{\sigma_1}
\AP_{\eta' 1}^{[\sigma_1]\dagger} \AP_{1 \eta}^{[\sigma_1]} =
\delta_{\eta' \eta}$ or $\sum_{\sigma_K} \AP_{\nu 1}^{[\sigma_K]}
\AP_{1 \nu'}^{[\sigma_K]\dagger } = \delta_{\nu \nu'}$, respectively.
In the NRG approach, all $A$-matrices naturally
satisfy \Eq{eq:orthonormality-left-to-right} [cf.
\Eq{eq:orthonormality}]. 

In the VMPS approach, it is convenient to ensure that during  
variational optimization sweeps, 
\Eq{eq:orthonormality-left-to-right} holds for all 
matrices to the left of the site, say $k$, currently being updated, 
and \Eq{eq:orthonormality-right-to-left} for all
matrices to its right. Thus, after optimizing the set 
of matrices $\AP^{[\sigma_k]}$ at site $k$, this set 
should be orthonormalized before moving
on to the next site, such that it satisfies
\Eq{eq:orthonormality-left-to-right} when sweeping from 
left to right (or \Eq{eq:orthonormality-right-to-left} 
when sweeping from right to left). 
This can be achieved using  singular value
decomposition [cf.\ \Eq{eq:SVD}]:
Arrange the matrix elements of the set of matrices
$\AP^{[\sigma_k]}$ into a rectangular matrix carrying only
two labels, with matrix elements ${\cal B}_{\bar \nu \eta} =
\AP^{[\sigma_k]}_{\nu \eta}$ (or ${\cal B}_{\nu \bar \eta} =
\AP^{[\sigma_k]}_{\nu \eta}$), by introducing a composite index $\bar \nu
= (\sigma_k, \nu)$ (or $\bar \eta = (\sigma_k, \eta)$).  Using singular
value decomposition [\Eq{eq:SVD}], write this new matrix as ${\cal B}
= \UUU \SSS \VVV^\dagger$.  Then rewrite the matrix product of two
neighboring $\AP$-matrices as $\AP^{[\sigma_k]} \AP^{[\sigma_{k+1}]} =
\tilde \AP^{[\sigma_k]} \tilde \AP^{[\sigma_{k+1}]}$ (or
$\AP^{[\sigma_{k-1}]} \AP^{[\sigma_k]} = \tilde \AP^{[\sigma_{k-1}]}
\tilde \AP^{[\sigma_k]}$), where the new matrices $\tilde \AP$ are
defined by
\begin{eqnarray}
  \tilde \AP^{[\sigma_k]}_{\nu \gamma} &  =  & \UUU_{\bar \nu \gamma} \; , 
  \quad   
  \tilde \AP^{[\sigma_{k+1}]}_{\gamma \delta}  =  
  (\SSS \VVV^\dagger \AP^{[\sigma_{k+1}]})_{\gamma\delta} \; , 
\label{eq:orthonormalize1}  
\\
 ({\rm or} \; \;
  \tilde \AP^{[\sigma_k]}_{\delta \eta}   & = &  \VVV^\dagger_{\delta
    \bar \eta} \; , \quad 
  \tilde \AP^{[\sigma_{k-1}]}_{\gamma\delta} 
  =  (\AP^{[\sigma_{k-1}]} \UUU \SSS)_{\gamma\delta}\; ).
\qqph
\label{eq:orthonormalize2}
\end{eqnarray}
The property $\UUU^\dagger \UUU = \mathone $
(or $\VVV^\dagger \VVV = \mathone $) ensures that the new
set of matrices $\tilde \AP^{[\sigma_k]}$ at site $k$ is orthonormal
according to \Eq{eq:orthonormality-left-to-right} (or
\Eq{eq:orthonormality-right-to-left}), as desired.  Now proceed to the
next site to the right (or left) and orthonormalize $ \tilde
\AP^{[\sigma_{k+1}]} $ (or $ \tilde \AP^{[\sigma_{k-1}]} $) in the same
manner, etc.

The above procedure can be used to orthonormalize the matrices of a
randomly generated matrix product state before starting VMPS sweeping.
Likewise, during VMPS sweeping, each newly optimized matrix can be
orthonormalized in the above fashion before moving on to optimize the
matrix of the next site. 

\subsection{Refolding}
\label{app:refolding}

This subsection describes how to refold an unfolded matrix product
state of the form 
\begin{eqnarray}
  |\Psi^{n}_{\nu\eta} \rangle_\unfolded =
  \sum_{\lbrace\bsigma^N \rbrace} \! 
  |\bsigma^n\rangle
  (\AP^{[\sigma_{n\downarrow}]} \!
  \dots \AP^{[\sigma_{0\downarrow}]}
  \AP^{[\sigma_{0\uparrow}]} \!
  \dots \AP^{[\sigma_{n\uparrow}]})_{\nu\eta} ,
  \nonumber \hspace{-0.5cm}  \phantom{.} 
  \\
\label{eq:DMRG_MPS_app}
\end{eqnarray}
shown schematically by sites $n\!\! \downarrow$ to $n\!\! \uparrow$ of
Fig.~\ref{fig:matrix-product-structure}(c).  Its two indices will be
treated as a composite index $\beta = (\nu,\eta)$ below.  The
variational matrix product state $|\Psi^N\rangle_\unfolded$ of
\Eq{eq:DMRG_MPS} discussed in the main text is a special case of
\Eq{eq:DMRG_MPS_app}, with $n=N$ and $\nu=\eta=1$ .  The goal is to
express \Eq{eq:DMRG_MPS_app} as a linear combination,
\begin{equation}
  \label{eq:unfolded=refoldedC}
  |\Psi^{n}_{\nu\eta} \rangle_\unfolded 
=  \sum_{\alpha} |\Psi^n_\alpha \rangle_\refolded \,
  C^n_{\alpha \beta} \,  ,
\end{equation}
($\beta = (\nu, \eta)$ is a composite index) 
 of an orthonormal set of ``refolded basis states'' of the form
of \Eq{eq:refolded_DMRG_MPS_basis}, 
\begin{eqnarray}
  |\Psi^n_\alpha\rangle_\refolded =
  \sum_{\lbrace\bsigma^n \rbrace}
  |\bsigma^n \rangle
  (\AP^{[\sigma_0]} \AP^{[\sigma_1]} \dots \AP^{[\sigma_n]})_{1 \alpha}
  \; , 
\label{eq:refolded_DMRG_MPS_app}
\end{eqnarray}
shown schematically by sites 0 to $n$ of
Fig.~\ref{fig:matrix-product-structure}(a).  
To this end, we procede
iteratively in $n$. We use singular value decomposition to iteratively
merge, for every pair of sites $n \! \! \downarrow$ and $n\! \!
\uparrow$ of the unfolded chain, the matrices
$\AP^{[\sigma_{n\downarrow}]}_{\nu \nu'}$ and
$\AP^{[\sigma_{n\uparrow}]}_{\eta'\eta}$ into a new set of matrices
$\AP^{[\sigma_n]}_{\alpha' \alpha}$ for site $n$ of the refolded
chain, thereby trading the indices
$\sigma_{n\downarrow},\sigma_{n\uparrow}$ and $\nu\eta$ of
Fig.~\ref{fig:matrix-product-structure}(c) for the indices $\sigma_n$
and $\alpha$ of Fig.~\ref{fig:matrix-product-structure}(a).  This is
to be done in such a way that the matrices $\AP^{[\sigma_{n}]}$ are
orthonormal in the sense of \Eq{eq:orthonormality}, and that for the
first few sites their dimensions increase in a way analogous to those
of the $A^{[\sigma_n]}$ matrices of NRG, starting from $1\times \dNRG$
at site $n=0$.

 For the first
iteration step, start with $n=0$, make a singular value decomposition
of the matrix product
\begin{equation}
  \label{eq:B0B0B0}
  ( \AP^{[\sigma_{0 \downarrow}]}
  \AP^{[\sigma_{0 \uparrow}]})_{\nu'\eta'}
  = (\UUU^{0} \SSS^{0} \VVV^{0\dagger} )_{\sigma_0 \beta'},   
\end{equation}
with  
$\beta' = (\nu',\eta')$, and use $\UUU^0$ to define a new set of
$\dNRG$ matrices $\AP^{[\sigma_0]} $ for site $0$ of the refolded
chain, with matrix elements $\AP^{[\sigma_0]}_{1 \alpha'} =
\UUU^{0}_{\sigma_0 \alpha'} $.  The $\AP^{[\sigma_0]} $ have
dimensions $1\times \dNRG$ (the dummy first index has just one value),
and are by construction orthonormal in the sense of
\Eq{eq:orthonormality}, since $\UUU^{0\dagger} \UUU^{0} = \mathone$.
Upon inserting \Eq{eq:B0B0B0} into \Eq{eq:DMRG_MPS_app}, the factor
$\UUU^{0}$ produces the first matrix factor $\AP^{[\sigma_0]}$ in the
refolded state (\ref{eq:refolded_DMRG_MPS_app}), thus completing the
first iteration step.  For the second iteration step, contract the
factors $\SSS^{0} \VVV^{0\dagger}$ with the factors $ \AP^{[\sigma_{1
    \downarrow}]}$ and $\AP^{[\sigma_{1 \uparrow}]}$ in
\Eq{eq:DMRG_MPS_app}, factorize the result as $\UUU^{1} \SSS^{1}
\VVV^{1\dagger}$ and use $ \UUU^{1}$ to construct new matrices
$\AP^{[\sigma_1]}$ for site $1$ of the refolded chain, etc.  To be
explicit, for general $n$, make a singular value decomposition of the
matrix product
\begin{equation}
  \label{eq:BnBnBn}
\sum_{\nu'\eta'}  \AP^{[\sigma_{n \downarrow}]}_{\nu \nu'} 
(\SSS^{(n-1)} \VVV^{(n-1)\dagger})_{\alpha' \beta'} 
  \AP^{[\sigma_{n \uparrow}]}_{\eta' \eta}
  = (\UUU^{n} \SSS^{n} \VVV^{n \dagger} )_{\bar \alpha \beta},   
\end{equation}
with composite indices $\bar \alpha = (\sigma_n, \alpha')$, $\sigma_n
= (\sigma_{n \downarrow}, \sigma_{n \uparrow})$, $\beta = (\nu, \eta)$
and $\beta' = (\nu', \eta')$.
Then use $\UUU^n$ to define a new set of orthonormal matrices
$\AP^{[\sigma_n]} $ for site $n$ of the refolded chain, with matrix
elements $\AP^{[\sigma_n]}_{ \alpha' \alpha} = \UUU^{n}_{\bar \alpha
  \alpha} $.  In this way one readily establishes that $
|\Psi^{n}_{\nu\eta} \rangle_\unfolded $ can be written in the form of
\Eq{eq:unfolded=refoldedC}, with 
$C^n_{\alpha \beta} = (\SSS^{n} \VVV^{n\dagger})_{\alpha \beta}$.

The dimensions of the matrices $\AP^{[\sigma_n]}$ initially grow by a
factor of $\dNRG$ with each iteration step, until their dimensions are
restricted by the number of possible values of the composite index
$\beta$, namely $\DDMRG^2_n$, with $\DDMRG_n$ given by
\Eq{eq:DDMRG_n}. Thus, the $\AP^{[\sigma_n]}$ have dimensions
$\DNRG_n^\refolded \times \DNRG_{n+1}^\refolded$, with
$\DNRG_{n}^\refolded = \min(\dNRG^{n}, \DDMRG^2_{n-1})$, which leads to
\Eq{eq:DNRG-refolded}.

\begin{figure}[tb]
 \centering \includegraphics[width=1.02\linewidth]{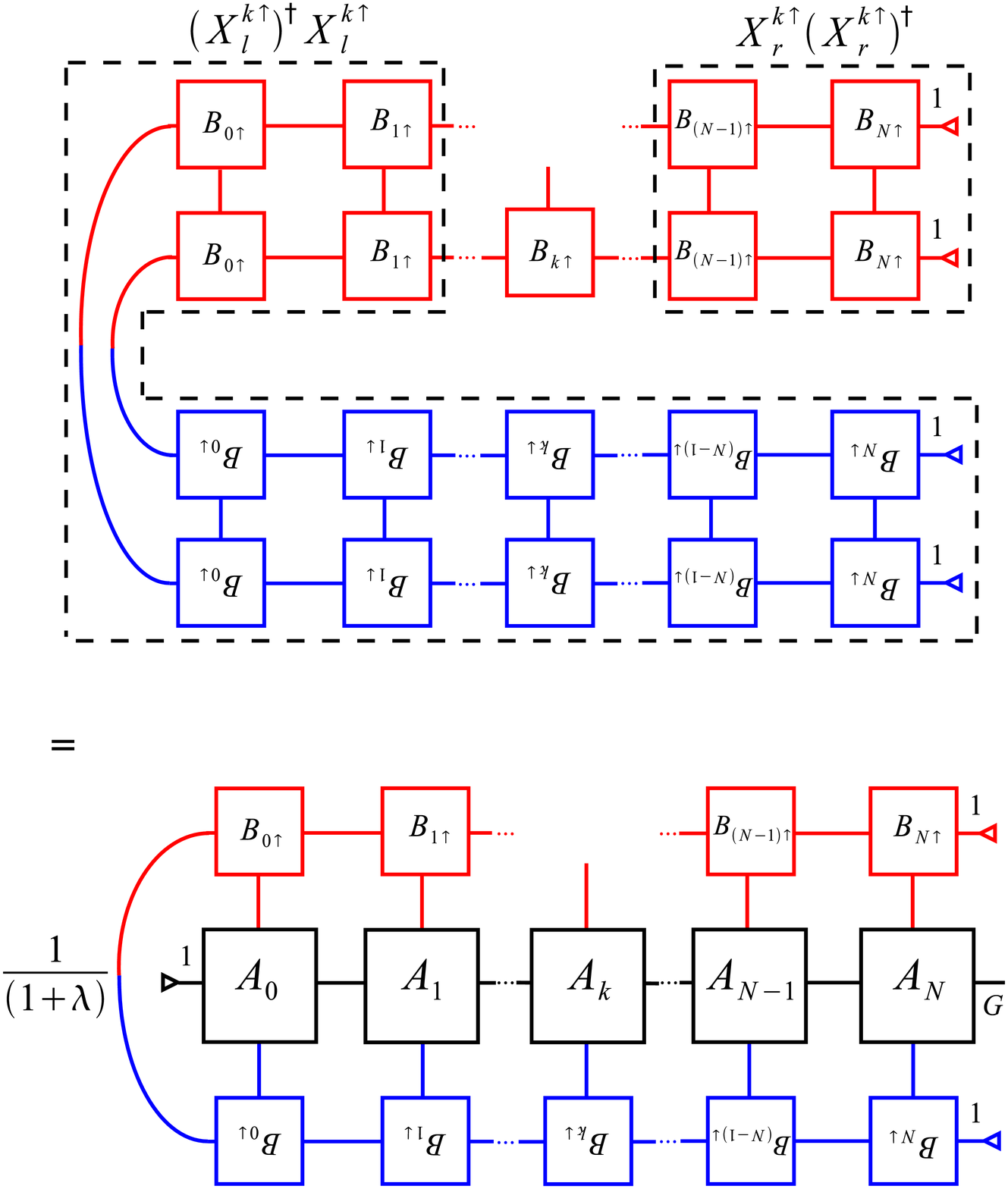} 
\caption{(Color online) Graphical representation of the 
variational equation  used for cloning, \Eqs{eq:optim_problem}
or (\ref{eq:determineBstar}), drawn for the case $\mu = \uparrow$,
and assuming all matrix elements to be real.
The upper part of the figure represents 
$\frac{1}{2} \frac{\partial }{\partial B^{[k \uparrow]}} 
{}_\cloned \langle G |  G \rangle_\cloned $; it 
simplifies to $B^{[k \uparrow]}$ [left hand side of \Eq{eq:determineBstar}]
upon realizing that the parts in dashed boxes represent
the left hand sides of \Eqs{eq:XlXl} and (\ref{eq:XrXr}),
and hence reduce to unity.}
\label{fig:clm}
\end{figure}

\subsection{Cloning}
\label{app:cloning}
This subsection gives some details of the 
cloning procedure of Section~\ref{sec:cloning}.
The goal is to solve the variational \Eq{eq:optim_problem},
which determines the $B$-matrices of the cloned state
$|G\rangle_\cloned$.  As described in the main text, this can be done 
by sweeping back and forth along the unfolded Wilson chain,
and updating one matrix at a time.

Let $k\mu$ label the site to be updated and write the cloned state, which
is assumed to be of the form (\ref{eq:DMRG_MPS}), as
\begin{equation}
  \label{eq:clonedXBX}
  |G\rangle_\cloned = 
\Bigl( X^{k\mu}_l \Bigr)_{1 \nu}
B^{[\sigma_{k\mu}]}_{\nu \nu'}
\Bigl( X^{k\mu}_r \Bigr)_{\nu' 1} \; . 
\end{equation}
Here we introduced the shorthands
\begin{subequations}
  \label{subeq:Xstrings}
\begin{eqnarray}
  \label{eq:Xl}
  \Bigl( X^{k\mu}_l \Bigr)_{1 \nu} & = & 
 \Bigl( B^{[\sigma_{N \downarrow}]} \dots 
B^{[\sigma_{k_l   \mu_l}]}  \Bigr)_{1 \nu} \; , 
\\ 
\label{eq:Xr}
  \Bigl( X^{k\mu}_r \Bigr)_{ \nu 1 } & = & 
\Bigl(B^{[\sigma_{k_r \mu_r}]} 
\dots 
 B^{[\sigma_{N \uparrow}]} \Bigr)_{\nu 1} 
\; , 
\end{eqnarray}
\end{subequations}
for the products of matrices standing before or after the one of
present interest in the unfolded Wilson chain, and the labels $k_l
\mu_l$ or $k_r \mu_r$ label the sites just before or after this site.
Moreover, assume that all the $B$-matrices in $X_l$ and $X_r$ have
been orthonormalized according to \Eq{eq:orthonormality-left-to-right}
or (\ref{eq:orthonormality-right-to-left}), respectively. (This can
always be ensured by suitably orthonormalizing each $B$-matrix after
updating it, see below.) These orthonormality relations immediately
imply similar ones for the matrix products just introduced:
\begin{subequations}
\label{subeq:Xorthonormal}  
\begin{eqnarray}
  \label{eq:XlXl}
  \sum_{\sigma_{N \downarrow}, \dots, \sigma_{k_l \mu_l}}
  \Bigl( X^{k\mu}_l \Bigr)^\dagger_{\nu 1}
  \Bigl( X^{k\mu}_l \Bigr)_{1 \nu'}   & = & \delta_{\nu \nu'} \; , 
\\
  \label{eq:XrXr}
  \sum_{\sigma_{k_r \mu_r} , \dots, \sigma_{N \uparrow}}
\Bigl( X^{k\mu}_r \Bigr)_{\nu 1}
  \Bigl( X^{k\mu}_r \Bigr)^\dagger_{1 \nu'}   
& = & \delta_{\nu \nu'} \; .  \qquad 
\end{eqnarray}
\end{subequations}
Thus, the norm of $|G \rangle_\cloned$ can be written as
\begin{eqnarray}
  \label{eq:normGcB}
{}_\cloned \langle G |  G \rangle_\cloned  = 
{ 1 \over {\cal N}} \sum_{\nu \nu'} 
B^{[\sigma_{k\mu}] \dagger}_{\nu' \nu}  
B^{[\sigma_{k\mu}]}_{\nu \nu'}  \; , 
\end{eqnarray}
where ${\cal N}$ is a normalization constant  
ensuring that the  norm equals unity. 

Using  \Eq{eq:normGcB}, the variational \Eq{eq:optim_problem}
readily reduces to
\begin{eqnarray}
  \label{eq:determineBstar}
  B^{[\sigma_{k\mu}] }_{\nu \nu'} &  = &
  \sum_{\lbrace\bsigma'^N \rbrace} 
  \frac{(A^{[\sigma_N]\dagger} \dots A^{[\sigma_0]\dagger})_{\ground
      1} }{1 + \lambda}    
  \Bigl( X^{k\mu}_l \Bigr)_{1 \nu}
  \Bigl( X^{k\mu}_r \Bigr)_{\nu' 1} \; , 
  \nonumber
  \\
\end{eqnarray}
where $\lbrace\bsigma'^N \rbrace$ denotes the local indices of all
sites except the index $\sigma_{k \mu}$ of site $k\mu$,
and we have assumed all $A$- and $B$-matrices to be purely real
(exploiting the time-reversal invariance of the present model). 
This equation
completely determines the new matrix $B^{[\sigma_{k \mu}]}$ in terms
of the $A$-matrices specifying the NRG input state $|G\rangle_\folded$
and the $B$-matrices of sites other than the present one, which had
been kept fixed during this variational step.

Having calculated $B^{[\sigma_{k \mu}]}$, it should be properly
orthonormalized, following the procedure of \Eq{eq:orthonormalize1} or
\Eq{eq:orthonormalize2}, depending on whether we are sweeping from
left to right or vice versa. In other words, use the singular value
decomposition ${\cal U} {\cal S} {\cal V}^\dagger$ of the new-found
matrix $B^{[\sigma_{k \mu}]}$, to transfer a factor ${\cal S} {\cal
  V}^\dagger$ or ${\cal U} {\cal S}$ onto its right or left neighbor,
respectively, and rescale this neighbor by an overall constant to
ensure that the new state $|G \rangle_\cloned $ is still normalized
to unity. This concludes the update of site $k \mu$. Now move on to
its neighbor, etc., and thus sweep back and forth through the unfolded
Wilson chain, until convergence is reached.


\begin{thebibliography}{10}
\bibitem{Wilson} K.G. Wilson, Rev. Mod. Phys. \textbf{47}, 773 (1975).

\bibitem{Krishna} H.R. Krishna-murthy, J.W. Wilkins, and K. G. Wilson,
Phys. Rev. B \textbf{21}, 1003 (1980).

\bibitem{Bulla} R. Bulla, T. Costi, and T. Pruschke, cond-mat/0701105.

\bibitem{AndersSchiller} 
F. B. Anders and A. Schiller, 
Phys.\ Rev.\ Lett.\ {\bf 95}, 196801 (2005);
Phys.\ Rev.\ B {\bf 74}, 245113 (2006). 

\bibitem{AndersPruschkePRB06}
R. Peters, T. Pruscke, and F. B. Anders, 
Phys.\ Rev.\ B {\bf 74}, 245114 (2006).

\bibitem{WV} A. Weichselbaum and J. von Delft, Phys. Rev. Lett. \textbf{99},
076402 (2007).

\bibitem{White} S. White, Phys. Rev. Lett. \textbf{69}, 2863 (1992).

\bibitem{White2} S. White, Phys. Rev. B \textbf{48}, 10345 (1993).

\bibitem{Schollwoeck} U. Schollwöck, Rev. Mod. Phys. \textbf{77},
259 (2005).

\bibitem{Ostlund_Rommer} S. Ostlund and S. Rommer, Phys. Rev. Lett.
\textbf{75}, 3537 (1995).

\bibitem{Dukelsky} J. Dukelsky \textit{et al.}, Europhys.Lett. \textbf{43},
457 (1998).

\bibitem{stoch_MPS} A. Honecker, I. Peschel, J. Stat. Phys. \textbf{88}
(1997).

\bibitem{AKLT+Fannes} M. Fannes, B. Nachtergaele and R. F. Werner,
Comm. Math. Phys. \textbf{144}, 443 (1992).


\bibitem{VPC} F. Verstraete, D. Porras and J.I. Cirac, Phys. Rev.
Lett. \textbf{93}, 227205 (2004).


\bibitem{VMPS} F. Verstraete, A. Weichselbaum, U. Schollwöck, J.
I. Cirac, and J. von Delft, cond-mat/0504305.

\bibitem{unpublished}
A. Weichselbaum (unpublished).

\bibitem{Raas} C. Raas, G.S. Uhrig and F.B. Anders, Phys. Rev. B
\textbf{69}, 041102(R) (2004) 

\bibitem{JW} P. Jordan and E.P. Wigner, Z. Phys. \textbf{47}, 631
(1928).



\bibitem{completebasis}
It has recently been shown \cite{AndersSchiller} that the discarded
states can be used to construct a complete
basis of many-body states spanning the full $d^{N+1}$-dimensional
Hilbert space of ${\cal H}_N$, and that this basis
can be used to greatly improve the accuracy
of the NRG-calculations of spectral functions.\cite{AndersPruschkePRB06,WV}


\bibitem{Muender}
A. Weichselbaum et al.\
to be published.

\bibitem{Holzner}
A. Holzner, A. Weichselbaum, J. von Delft,
to be published. 

\bibitem{reducedDM}
M.A. Nielsen and I.L. Chuang,
\emph{Quantum Computation and Quantum Information}
Cambridge University Press (2000).

\bibitem{PustilnikGlazmanJPhysCM04} 
M. Pustilnik and L. Glazman, J. Phys.: Condens. Matter {\bf 16}, R513 (2004).

\bibitem{TothZarandArXiV08}
A.I. T\'{o}th, C.P. Moca, O. Legeza, G. Zar\'{a}nd,
arXiv:0802.4332.


\bibitem{McCouloghArXiV08}
I.P. McCulloch, J. Stat.\ Mech.\ P10014 (2007).

\end{thebibliography}
\end{document}